\documentclass[aps,prb,reprint,superscriptaddress]{revtex4-2}
\usepackage{graphicx}
\usepackage{hyperref}
\usepackage{amssymb}
\usepackage{mathrsfs}
\usepackage{amsmath}

\hypersetup{
	colorlinks=true,
	anchorcolor=blue,
	linkcolor=blue,
    citecolor=blue,
    urlcolor=blue
}

\begin{document}

\title{Andreev reflection in normal metal/charge-4e superconductor junctions}

\author{Yi-Xin Dai}
\affiliation{International Center for Quantum Materials, School of Physics, Peking University, Beijing 100871, China}
\affiliation{Beijing Academy of Quantum Information Sciences, West Bld.\#3,No.10 Xibeiwang East Rd., Haidian District, Beijing 100193, China}

\author{Qing-Feng Sun}
\email{sunqf@pku.edu.cn}
\affiliation{International Center for Quantum Materials, School of Physics, Peking University, Beijing 100871, China}
\affiliation{Beijing Academy of Quantum Information Sciences, West Bld.\#3,No.10 Xibeiwang East Rd., Haidian District, Beijing 100193, China}
\affiliation{Hefei National Laboratory, Hefei 230088, China}

\date{\today}

\begin{abstract}
We investigate the Andreev reflection in a normal metal/charge-$4e$ superconductor junction. 
Compare with the electron-hole conversion in normal charge-$2e$ superconductors, here four electrons participate  simultaneously, enriching the possibility of conversion ways. 
Using nonequilibrium Green's function method, we obtain a four-particle-type Laudauer-B\"uttiker formula with generalized charge-$4e$ anomalous Green's function to describe it.
We then calculate and clarify the behavior of the Andreev coefficient with various incident energy and show the conductance contributed by it.
Our research makes up the blank of research for transport property of charge-$4e$ superconductors and can be served as hallmarks for future experimental verifications.

\end{abstract}

\maketitle

\section{\label{intro}Introduction}
Charge-$4e$ superconductivity (charge-$4e$ SC) names the condensate of the quartets of electrons.
Such electron pairs can carry a charge of $4e$, which differs from the normal charge-$2e$ Cooper pairs described by the Bardeen-Cooper-Schrieffer theory\cite{bardeen_Theory_1957}.
It has been proposed in a wide range of fields\cite{ropke_FourParticle_1998,wu_Competing_2005,aligia_Quartet_2005,sogo_Manybody_2010,moon_Skyrmions_2012,fernandes_Charge_2021,jian_Charge_2021,hecker_Cascade_2023}.
Especially in the theory of intertwined orders in cuprate superconductors, the charge-$4e$ SC is described as a vestigial phase in the thermal melting of pair-density-wave order\cite{berg_Charge4e_2009,fradkin_Colloquium_2015}.
Recent progress in experiments makes charge-$4e$ SC more attractive\cite{grinenko_State_2021,ge_Discovery_2022}, due to the possible evidence for charge-$4e$ and even charge-$6e$ pairs in kagome supercondcutors.
Although theorists have proposed possible mechanism for charge-$4e/6e$ pairs\cite{agterberg_Conventional_2011,zhou_Chern_2022,garaud_Effective_2022},
it still requires more evidence to confirming the existence of charge-$4e$ SC in experiments.
Therefore, it is quite necessary to investigate the transport properties of charge-$4e$ SC in the hope of giving more hallmarks to distinguish it from the perspective of transport phenomena.

The former researches are mainly focus on the equilibrium properties of charge-$4e$ SC\cite{herland_Phase_2010,radzihovsky_Fluctuations_2011,radzihovsky_Fluctuations_2011,barci_Role_2011,moon_Skyrmions_2012,wang_Braiding_2016,fernandes_Charge_2021,jian_Charge_2021,khalaf_Symmetry_2022}.
By applying the theory of vestigial order and the Ginzburg-Landau method, they showed the appearance of charge-$4e$ SC order and discussed the phase competition between it and other orders, such as pair-density-wave and nematic orders\cite{barci_Role_2011,radzihovsky_Fluctuations_2011,hecker_Cascade_2023,curtis_Stabilizing_2023,varma_Extended_2023}.
While those studies were phenomenological in general, some works began to pay attention to the microscopic mechanics of forming four-electron pairs\cite{tarasewicz_Extension_2006,jiang_Charge4e_2017,gnezdilov_Solvable_2022,li_Charge_2022}.
Such as the quantum Monte Carlo simulations on the real space mean-field Hamiltonian, in which four electrons on the neighboring sites can form a pair,
showed a chemical potential controlled phase transition between charge-$4e$ SC and charge-$2e$ SC\cite{jiang_Charge4e_2017}.
And the solvable Sachdev-Ye-Kitaev model described a charge-$4e$ supercondcutor with gapless ground state\cite{gnezdilov_Solvable_2022}.
Besides, similar to the BCS wavefunction for charge-$2e$ SC, a charge-$4e$ wavefunction with its corresponding mean-field Hamiltonian have also been proposed\cite{li_Charge_2022}.
While this Hamiltonian describes a gapped charge-$4e$ SC phase, it can also be seen as a natural extension from the theory of intertwined orders.

To study the transport with superconductors, one often encounters the interface between leads and superconductors.
It is well known that an incident electron can be reflected as a hole at the interface between a normal metal and a superconductor, namely the Andreev reflection.
This process can convert the normal current into the supercurrent, governing the conductance of the interface below the superconducting gap.
Moreover, as the incident electron and reflected hole can be regulated by both sides of the junction, the Andreev reflection is often used to reveal the novel properties of materials either in the normal side\cite{beenakker_Specular_2006,titov_InteractionInduced_2006,hou_Crossed_2016,hou_Double_2017} or in the superconducting side\cite{lv_Magnetoanisotropic_2018,cuevas_Shot_2002,sukhachov_Andreev_2023,lewandowski_Andreev_2023}.
Therefore, in order to understand the transport with charge-$4e$ SC, it is natural to consider the possible Andreev reflection happening in the interface between a normal metal and a charge-$4e$ superconductor.
At this point, the conventional picture involving two participating electrons may be inapplicable due to the electron quartet, which introduces more freedom as well as the complexity.

In this paper, we investigate the charge-$4e$ Andreev reflection (charge-$4e$ AR) in a normal metal/charge-$4e$ superconductor junction.
Due to the characteristics of quartet condensation, the Andreev reflection here involves four particles, enriching the possibility of conversion ways.
Using nonequilibrium Green's function method, we derive a four-particle-type Laudauer-B\"uttiker formula with generalized charge-$4e$ anomalous Green's function to describe the Andreev reflection process.
We then calculate and clarify the behavior of charge-$4e$ AR with various
incident energy and show the conductance contributed by it.
Our results provide new formula and pictures to describe the charge-$4e$ AR,
which can enrich the understanding of the transport with charge-$4e$ SC
and give guidance for the future experimental verifications.

The rest of the paper is as follows: In Sec.\ref{model}, we give the Hamiltonian of the normal metal/charge-$4e$ superconductor junction and derive the formula for the transport process.
In Sec.\ref{per} and\ \ref{nonper}, we calculate the Andreev coefficient using perturbation and nonperturbation methods, which characterize the Andreev reflection from different perspectives while maintaining some consistency.
We then give the whole picture of charge-$4e$ AR in Sec.\ref{ar}
and further consider the conductance in Sec.\ref{cond}.
Finally, a brief summary is presented in Sec.\ref{diss}.

\section{\label{model}Model and formula}

We consider a normal metal/charge-$4e$ superconductor junction, which is described by the following Hamiltonian $H_{tot}=H_L + H_R + H_C \nonumber$, with
\begin{eqnarray}
    &H_L& = \sum_{i,k,\sigma} \epsilon_{L,ik}\
    a^\dagger_{ik\sigma} a_{ik\sigma},
    \nonumber\\
    &H_R& = \sum_{i,k,\sigma} \epsilon_{R,ik}\
    c^\dagger_{ik\sigma} c_{ik\sigma}
    + H_{sc},
    \nonumber\\
    &H_C& = \sum_{i,k,k',\sigma} t_{Lk,Rk'}\
    a^\dagger_{ik\sigma} c_{ik'\sigma} + h.c. ,
\end{eqnarray}
where $H_L$, $H_R$, and $H_C$ represent the Hamiltonians of left lead (normal metal),
right lead (charge-$4e$ superconductor) and their coupling, respectively.
$a_{ik\sigma}$ $(c_{ik\sigma})$ are electron annihilation operators in the left (right) lead, with $k$ denoting the momentum and $\sigma=\uparrow, \downarrow$ denoting the electron spins.
Besides, $\epsilon_{L/R,ik}$ and $t_{Lk,Rk'}$ are the energy dispersion and hopping strength.
We set $\epsilon_{L/R,ik}=\epsilon_k$ and $t_{Lk,Rk'}=t_c$ below for convenience.
$H_{sc}$ described the superconducting interaction.
Adopting the charge-$4e$ mean-field Hamiltonian in Ref.\cite{li_Charge_2022} and introducing the index $i=1,2$ to fulfill the Pauli exclusion principle, we have
\begin{eqnarray}
    &H^{4e}_{sc}& = \Delta\sum_{k}(
        c^\dagger_{1k\uparrow}c^\dagger_{1-k\downarrow}c^\dagger_{2k\uparrow}c^\dagger_{2-k\downarrow}+h.c.
    ),
\end{eqnarray}
with $\Delta$ being the superconducting pairing potential.
To compare with the charge-$2e$ SC, we also write the charge-$2e$ Hamiltonian $H^{2e}_{sc} = \Delta\sum_{k}(c^\dagger_{1k\uparrow}c^\dagger_{1-k\downarrow}+h.c.)$ with the index $i=1$ only.

Due to the symmetry of $H_{tot}$, the total current is four times that of a single channel current.
We thus consider the current flowing from channel 1$\uparrow$ via the time derivative of electron number operator $N_{1\uparrow,L}=\sum_k a^\dagger_{1k\uparrow} a_{1k\uparrow}$\cite{sun_Quantum_2009}
\begin{equation}\label{eq_current}
    I_{1\uparrow} =
    -e\langle\dot{N}_{1\uparrow,L}  \rangle
    = \frac{ie}{h} \int \mathrm{d}\omega_1
    \Gamma^L \left[
		f_{1\uparrow}^e G^>_{1\uparrow}
        + \bar{f}_{1\uparrow}^e G^<_{1\uparrow}
	\right] ,
\end{equation}
where $\Gamma^L=2\pi\rho |t_c|^2$ with $\rho$ being the density of states (DOS) in the normal lead.
$\bar{f}^e_{1\uparrow} = 1-f^e_{1\uparrow}$ and
$ f^e_{1\uparrow} = f(\omega_1-eV) =1/\left[e^{(\omega_1-eV)/k_B \mathcal{T}}+1\right]$ is the Fermi distribution function of channel 1$\uparrow$ in electron-type with the bias voltage $V$ and temperature $\mathcal{T}$.
The less and greater Green's functions $G^\lessgtr_{1\uparrow}(\omega_1)$ are the Fourier transforms of $G^\lessgtr_{1\uparrow}(t,0)$, which are defined as $G^<_{1\uparrow}(t,0)= i\sum_{kk'} \langle c_{1k'\uparrow}^\dagger(0) c_{1k\uparrow}(t) \rangle$ and $G^>_{1\uparrow}(t,0)= -i\sum_{kk'} \langle c_{1k\uparrow}(t) c_{1k'\uparrow}^\dagger(0)\rangle$.
Eq. (\ref{eq_current}) is a simple transform of the usual Landauer formula\cite{meir_Landauer_1992} and contains all the possible transport processes of an interacting system.

Before starting our procedure in calculating charge-$4e$ AR, we first briefly review the Andreev reflection in charge-$2e$ SC\@.
In order to describe charge-$2e$ AR, we need to explicitly insert the coupling term of channel 1$\downarrow$ into the nonequilibrium Green's function $G_{1\uparrow,k_1k_1'}(\tau_1,\tau_1') = -i\langle c_{1k_1\uparrow}(\tau_1) c^\dagger_{1k_1'\uparrow}(\tau_1')\rangle^c$,
with superscript `c' denoting the complex contour time order and the time $\tau_1$ and $\tau_1'$ being defined on the contour\cite{sun_Heat_2007},
to do the perturbation expansion.
This would introduce the nonequilibrium anomalous Green's functions $F_{k_1k_2}(\tau_1,\tau_2)=-i\langle c_{1k_1\uparrow}(\tau_1) c_{1-k_2\downarrow}(\tau_2)\rangle^c$ and $F^\dagger_{k_2'k_1'}(\tau_2',\tau_1')=-i\langle c^\dagger_{1-k_2'\downarrow}(\tau_2') c^\dagger_{1k_1'\uparrow}(\tau_1') \rangle^c$.
After applying the analytic continuation and Fourier transform, we derive the part of the $G^\lessgtr_{1\uparrow}(\omega_1)$ relating to the process of Andreev reflection (denoted with subscript `A') as
\begin{eqnarray}\label{eq_g2e}
    G^>_{1\uparrow,A}(\omega_1)
    &=& F^r(\omega_1)
    \left[ -i \bar{f}^h_{1\downarrow} \Gamma^L \right]
    F^{\dagger,a}(\omega_1) ,
\nonumber\\
    G^<_{1\uparrow,A}(\omega_1)
    &=& F^r(\omega_1)
    \left[ i f^h_{1\downarrow} \Gamma^L \right]
    F^{\dagger,a}(\omega_1) ,
\end{eqnarray}
where $\bar{f}^h_{1\downarrow} = 1-f^h_{1\downarrow}$ and $f^h_{1\downarrow} = f(\omega_1+eV)$ is the Fermi distribution function of channel 1$\downarrow$ in hole-type.
$F^r(\omega_1)$ and $F^{\dagger,a}(\omega_1)$ are the Fourier transforms of $\sum_{k_1k_2} F^r_{k_1k_2}(t,0)$ and $\sum_{k_2'k_1'} F^{\dagger,a}_{k_2'k_1'}(t,0)$.
The retarded (advanced) Green's functions can be obtained from the analytic continuation
\begin{eqnarray}\label{eq_fr2e}
    F^r(t_1,t_2)
    &=& F (t_1^{s},t_2^+)
    - F (t_1^{s},t_2^-) ,
    \nonumber\\
    F^{\dagger,a} (t_2,t_1)
    &=& F^{\dagger} (t_2^+,t_1^{s})
    - F^{\dagger} (t_2^-,t_1^{s}) ,
\end{eqnarray}
with $s=+$ or $-$.
The arbitrary choosing of $s=\pm$ branch of the complex contour shows the causality\cite{jakobs_Properties_2010}, that $t_1>t_2$ in $F^r(t_1,t_2)$ and $F^{\dagger,a}(t_2,t_1)$.
Substituted Eq. (\ref{eq_g2e}) into Eq. (\ref{eq_current}), we obtain the current contributed by the Andreev reflection
\begin{equation}\label{eq_ja_2e}
    I_{1\uparrow,A}
    = \frac{e}{h} \int \mathrm{d}\omega_1
    (f^e_{1\uparrow}\bar{f}^h_{1\downarrow} - \bar{f}^e_{1\uparrow} f^h_{1\downarrow})\ T_A ,
\end{equation}
with $T_A(\omega_1) = \Gamma^L F^r \Gamma^L F^{\dagger,a}$ being the Andreev reflection coefficient.
Notice that $f^e_{1\uparrow}\bar{f}^h_{1\downarrow} - \bar{f}^e_{1\uparrow} f^h_{1\downarrow} = f^e_{1\uparrow}-f^h_{1\downarrow}$,
so the Andreev reflection current in Eq. (\ref{eq_ja_2e}) is identical with that in the literature\cite{cuevas_Hamiltonian_1996,sun_Resonant_1999,sun_Quantum_2009}.
For completeness, we leave the detailed derivation and the formula for other processes contributed to the total current in Appendix~\ref{tran_2e}.

It is then straightforward to generalize the charge-$2e$ AR to the charge-$4e$ AR\@.
By inserting the coupling terms of other three channels,
we introduce the nonequilibrium charge-$4e$ anomalous Green's function
\begin{eqnarray}\label{eq_f4e}
    &&F_{k_1k_2k_3k_4}
    (\tau_1,\tau_2,\tau_3,\tau_4)
    \nonumber\\
    &&\quad= -i\langle
        c_{1k_1\uparrow}(\tau_1)
        c_{1-k_2\downarrow}(\tau_2)
        c_{2k_3\uparrow}(\tau_3)
        c_{2-k_4\downarrow}(\tau_4)
    \rangle^c  ,
    \nonumber\\
    &&F^{\dagger}_{k_4k_3k_2k_1}
    (\tau_4,\tau_3,\tau_2,\tau_1)
    \nonumber\\
    &&\quad = -i\langle
        c^\dagger_{2-k_4\downarrow}(\tau_4)
        c^\dagger_{2k_3\uparrow}(\tau_3)
        c^\dagger_{1-k_2\downarrow}(\tau_2)
        c^\dagger_{1k_1\uparrow}(\tau_1)
    \rangle^c .
\end{eqnarray}
After applying the analytic continuation and the Fourier transform, we obtain
the part of the $G^\lessgtr_{1\uparrow}(\omega_1)$ relating to the process of charge-4$e$ AR as
\begin{eqnarray}\label{eq_g4e}
    &&G^>_{1\uparrow,A} (\omega_1)
\nonumber\\
    &&= -i\int_{234}
    F^r\left[
        \bar{f}^h_{1\downarrow}(\omega_2)
        \bar{f}^h_{2\uparrow}(\omega_3)
        \bar{f}^h_{2\downarrow}(\omega_4)
        {(\Gamma^L)}^3
    \right] F^{\dagger,a} ,
\nonumber\\
    &&G^<_{1\uparrow,A} (\omega_1)
\nonumber\\
    &&= i\int_{234}
    F^r \left[
        f^h_{1\downarrow}(\omega_2)
        f^h_{2\uparrow}(\omega_3)
        f^h_{2\downarrow}(\omega_4)
        {(\Gamma^L)}^3
    \right] F^{\dagger,a},
\end{eqnarray}
where $\bar{f}^h_{i\sigma}(\omega) = 1-f^h_{i\sigma}(\omega)$ and $f^h_{i\sigma}(\omega)=f(\omega+eV)$ is the Fermi distribution function of channel $i\sigma$ in hole-type.
The integral $\int_{234}$ represents the energy convolution $\int \mathrm{d}\omega_2 \mathrm{d}\omega_3 \mathrm{d}\omega_4 \delta(\omega_1-\omega_2-\omega_3-\omega_4)/4\pi^2$, constrainting the total energy of the three reflected holes to form four-electron pairs at the Fermi surface.
$F^r=F^r(\omega_2,\omega_3,\omega_4)$ and $F^{\dagger,a}=F^{\dagger,a}(\omega_4,\omega_3,\omega_2)$ are the Fourier transforms of $\sum_{k_i} F^r_{k_1k_2k_3k_4}(0,t_2,t_3,t_4)$ and $\sum_{k'_j}F^{\dagger,a}_{k'_4k'_3k'_2k'_1}(t_4,t_3,t_2,0)$, respectively.
We have the generalized retarded (advanced) Green's functions (omitting the momentum index for simplification)
\begin{eqnarray}\label{eq_fr4e}
    F^r(t_1,t_2,t_3,t_4)
    &&= \sum_{s_2s_3s_4} {(-1)}^P
    F(t^\pm_1,t^{s_2}_2,t^{s_3}_3,t^{s_4}_4) ,
    \nonumber\\
     F^{\dagger,a}(t_4,t_3,t_2,t_1)
    && = \sum_{s_2s_3s_4} {(-1)}^P
    F^{\dagger}(t^{s_4}_4,t^{s_3}_3,t^{s_2}_2,t^\pm_1) ,
\end{eqnarray}
with $s_{2,3,4}=\pm$ being the branch index and $P$ being the total number of `$-$' branch among them.
Note that the results in Eq. (\ref{eq_fr4e}) are the same regardless of $t_1$ either on
$+$ branch or on $-$ branch.
This arbitrary choosing of the branch of $t_1$ is protected by the causality that $t_1>t_2,t_3,t_4$.

Substituting Eq. (\ref{eq_g4e}) into Eq. (\ref{eq_current}) leads to a four-particle Landauer formula
\begin{equation}\label{eq_4e}
    I_{1\uparrow,A}
    = \frac{e}{h} \int_{1234}
    (
        f^e_{1\uparrow}
        \bar{f}^h_{1\downarrow}
        \bar{f}^h_{2\uparrow}
        \bar{f}^h_{2\downarrow}
        -
        \bar{f}^e_{1\uparrow}
        f^h_{1\downarrow}
        f^h_{2\uparrow}
        f^h_{2\downarrow}
    )\ T_A ,
\end{equation}
with $\int_{1234}$ being $\int \mathrm{d}\omega_1 \mathrm{d}\omega_2 \mathrm{d}\omega_3 \mathrm{d}\omega_4 \delta(\omega_1-\omega_2-\omega_3-\omega_4)/4\pi^2$ and $T_A(\omega_2,\omega_3,\omega_4) = \Gamma^L F^r {(\Gamma^L)}^3 F^{\dagger,a}$ being the charge-$4e$ Andreev reflection coefficient.
This is one of the central results of this work.
It describes the charge-$4e$ Andreev reflection process, where one incident electron can be reflected to three holes at the interface between normal metal and charge-$4e$ supercondcutor (shown in Fig.\ref{p_4e}), and can be easily extended to multi-electron Andreev reflection.
Note that since the Fermi distribution functions of electron and hole can be transformed by $\bar{f}^h_{i\sigma}(\omega^h)=f^e_{i\sigma}(\omega^e)$ with $\omega^e=-\omega^h$, the number of incident electrons and reflected holes just depends on the way we treat each channel as electron-type or hole-type.
We thus can rewrite Eq. (\ref{eq_4e}) as
\begin{equation}
    I_{1\uparrow,A}
    = \frac{e}{h} \int'_{1234}
    (
        f^{e}f^{e}f^{e}f^{e}
        -
        \bar{f}^{e} \bar{f}^{e}
        \bar{f}^{e} \bar{f}^{e}
    )\ T _A ,
\end{equation}
omitting the channel index $i\sigma$.
Here $\int'_{1234}$ denotes the energy integral $\int \mathrm{d}\omega^e_1 \mathrm{d}\omega^e_2 \mathrm{d}\omega^e_3 \mathrm{d}\omega^e_4 \delta(\omega^e_1+\omega^e_2+\omega^e_3+\omega^e_4)/4\pi^2$ with $\omega^e_1=\omega_1, \omega^e_{2,3,4}=-\omega_{2,3,4}$ unifying the energy variables in electron-type.

We then consider the conductance contributed by the charge-$4e$ AR,
which can be obtained from $G_A = \partial I_{1\uparrow,A}/\partial V$.
\begin{equation}\label{eq_ga}
    G_{A}
    = \frac{e^2}{h} \sum_{i=1}^4 \int'_{1234-i}
    (
        f^{e}f^{e}f^{e}
        +
        \bar{f}^{e}\bar{f}^{e}\bar{f}^{e}
    )\ T_A \Big|_{\omega^e_i=eV} ,
\end{equation}
assuming that $T_A$ is independent of the voltage $V$.
Here the integral $\int'_{1234-i}$ is over other three energy variables while keeping $\omega^e_i=eV$.
For completeness, we leave in Appendix~\ref{tran_4e} the detailed derivation as well as the transport formula for other processes and their contributions to the total conductance.
Recall that in charge-$2e$ SC, $G_A$ is simply proportional to $T_A$, the integral nature here would make the conductance of charge-$4e$ SC quite different from that of charge-$2e$ SC\@.

\section{\label{per}Perturbation in Superconductivity: lowest-order expansion}

It is challenging to solve the charge-$4e$ anomalous Green's functions $F,F^\dagger$ and give an exact form of $T_A$.
Therefore, we begin our analysis by considering some specific conditions where perturbation theory can take effect.

We first consider to expand the Andreev reflection coefficient to the lowest order of charge-$4e$ superconducting pairing potential $\Delta$.
Due to the many-body correlation nature of charge-$4e$ SC, in theoretical description, one usually has no particular reason to expect the charge-$4e$ pairing potential to be weak enough to apply a perturbation theory\cite{gnezdilov_Solvable_2022}.
However, such challenge can be avoided in transport problems, as we investigate the state that the energy of incident electrons are much greater than the superconducting gap.
Compare with the injecting energy, the pairing potential can now be seen as a small quantity, which makes the perturbation effective.

Therefore, we can expand the charge-$4e$ anomalous Green's function to the lowest order of $\Delta$ and obtain the Andreev reflection coefficient up to $\Delta^2$ order (here we also give the result of charge-$2e$ SC for comparation),
\begin{eqnarray}\label{eq_wd}
    &&T_{A,4e}^{(1)} = 4z^4\cdot\frac{{(\Delta/\pi\rho)}^2}{
        {(\omega_1+\omega_2)}^2
        {(\omega_1+\omega_3)}^2
        {(\omega_1+\omega_4)}^2
        } ,
\nonumber\\
    &&T_{A,2e}^{(1)} = 4z^2\cdot\frac{\Delta^2}{
        {(\omega_1+\omega_2)}^2
        } ,
\end{eqnarray}
with $z = 2\pi^2\rho^2 |t_c|^2/{(1+\pi^2\rho^2 |t_c|^2)}^2$ measuring the influence of coupling.
We have energy constraint $\omega_2=\omega_1$ for charge-$2e$ SC
and $\omega_2+\omega_3+\omega_4=\omega_1$ for charge-$4e$ SC, showing that the total energy carried by the reflected holes is equal to that of the incident electron.
A brief discussion for other processes is given in Appendix~\ref{per_other} for completeness.

It is now clear to see that the perturbation can take effect only when the denominators in Eq. (\ref{eq_wd}) are much greater than the numerators.
In charge-$2e$ SC, the energy constraint preserves that we just require the incident electron to be far away from the gap.
However, in charge-$4e$ SC, the loose constraint requires not only the incident electron but also the reflected holes should be far away from the gap.
The perturbation would also break down at some diverging points as some reflected holes carry the opposite energy of the incident electron.
However, these divergence can be compensated by considering the higher order contributions.
The maximum of $T^{(1)}_{A,4e}$ would happens when each reflected hole carries one-third of the energy of the incident electron.

Besides, we can see from Eq. (\ref{eq_wd}) that the structure of $T^{(1)}_{A,4e}$ is similar to the $T_{A,2e}^{(1)}$.
Since $T^{(1)}_{A,4e}\propto E^{-6}$ and $T_{A,2e}^{(1)}\propto E^{-2}$
with $E$ the incident energy, we find that their contributions to the conductance $G_A$ possess different energy dependence.
Meanwhile, the square of coupling from $z^2$ to $z^4$ makes $T^{(1)}_{A,4e}$ change faster than $T_{A,2e}^{(1)}$ as $z$ varies, which means that the conductance of charge-$4e$ SC is more sensitive to the interface barriers than that of charge-$2e$ SC\@.

\section{\label{nonper}Nonperturbation in Superconductivity: the equation of motion method}

As the incident (reflected) energy stays below or near the gap, the break down of the perturbation expanded by superconducting pairing potential $\Delta$
implies that we should consider higher orders of $\Delta$ into the anomalous Green's function.
One straightforward way is to expand the Andreev reflection coefficient to the lowest order of coupling $|t_c|^2$,
which makes the nonequilibrium charge-$4e$ anomalous Green's function $F_{k_1k_2k_3k_4}(0,t_2,t_3,t_4)$ regress to the equilibrium charge-$4e$ anomalous Green's function $\delta_{kk_1}\delta_{kk_2}\delta_{kk_3}\delta_{kk_4}F_k(0,t_2,t_3,t_4)$.
As the equilibrium Green's function of $H_R$ can be solved exactly\cite{li_Charge_2022},
for the convenience of future promotion to the weak coupling case, we introduce the equation of motion (EOM) for retarded charge-$4e$ anomalous Green's functions.

We first solve the equilibrium charge-$4e$ anomalous Green's functions of $H_R$
at $t_c=0$.
By using Eq. (\ref{eq_fr4e}) and the identity
\begin{eqnarray}
    1=\sum_{(xyz)}
    \theta(0-t_x) \theta(t_x-t_y) \theta(t_y-t_z) ,
\end{eqnarray}
with $(xyz)$ the permutation of $(234)$, we rewrite the retarded charge-$4e$ anomalous Green's function
\begin{eqnarray}\label{eq_fr}
    &&F^r_k(0,t_2,t_3,t_4)
\nonumber\\
&&\quad
    = \sum_{(xyz)}
    -i \mathscr{F}(
        \tilde{c}_{1k}(0),\tilde{c}_{xk}(t_x),
        \tilde{c}_{yk}(t_y),\tilde{c}_{zk}(t_z)
    )
\nonumber\\
&&\quad
    = \sum_{(xyz)} -i P(xyz)
    \theta(0-t_x) \theta(t_x-t_y) \theta(t_y-t_z)
\nonumber\\
&&\qquad
    \{[\{
        \tilde{c}_{1k}(0), \tilde{c}_{xk}(t_x)
    \}, \tilde{c}_{yk}(t_y)
    ], \tilde{c}_{zk}(t_z) \} ,
\end{eqnarray}
where $P(xyz)$ is the sign of the permutation $(xyz)$.
$\tilde{c}_{1k},\tilde{c}_{2k},\tilde{c}_{3k},\tilde{c}_{4k}$ represent $c_{1k\uparrow},c_{1-k\downarrow},c_{2k\uparrow},c_{2-k\downarrow}$, respectively.
Using the Fourier transform, we have
\begin{eqnarray}
    F^r_k(\omega_2,\omega_3,\omega_4)
    &&= \sum_{(xyz)}
    -i \mathscr{F}(\omega_x, \omega_y, \omega_z) .
\end{eqnarray}
The EOM for $F^r_k(\mathscr{F})$ can then be obtained by considering the time derivative, with detailed derivation shown in Appendix~\ref{eom_f}.

Recall the EOM for charge-$2e$ anomalous Green's function, we note that the closure of the EOM relies on the commutation relation
\begin{eqnarray}
    [c_{1k\sigma},H^{2e}_{sc}]
    = \Delta c^\dagger_{1-k-\sigma}
    \ ,\
    [c^\dagger_{1-k-\sigma},H^{2e}_{sc}]
    = \Delta c_{1k\sigma} ,
\end{eqnarray}
which also shows the particle-hole symmetry in charge-$2e$ SC\@.
Similarly, for charge-$4e$ SC, we have commutation relation
\begin{eqnarray}\label{eq_com}
    &&[c_{1k\uparrow}, H^{4e}_{sc}]
    = \Delta\ d^\dagger_{1k\uparrow} ,
\nonumber \\
    &&[ d^\dagger_{1k\uparrow}, H^{4e}_{sc}]
    = \Delta\ \xi_{1k\uparrow} c_{1k\uparrow} ,
\nonumber \\
    &&[\xi_{1k\uparrow}c_{1k\uparrow}, H^{4e}_{sc}]
    = \Delta\ d^\dagger_{1k\uparrow} ,
\end{eqnarray}
guaranteeing the closure of the EOM\@.
Here $d^\dagger_{1k\uparrow} = c^\dagger_{1-k\downarrow} c^\dagger_{2k\uparrow}c^\dagger_{2-k\downarrow}$ is the operator for charge-$3e$ particles.
$\xi_{1k\uparrow} = n_{1-k\downarrow} n_{2k\uparrow} n_{2-k\downarrow} + \bar{n}_{1-k\downarrow} \bar{n}_{2k\uparrow} \bar{n}_{2-k\downarrow}$ with $\bar{n}=1-n$ shows the effect of occupation numbers to the excitations.
Therefore, Eq. (\ref{eq_com}) expresses the particle-hole symmetry in charge-$4e$ SC, where a charge-$3e$ hole (particle) can be converted into a charge-$e$ particle (hole) combining with its environment (described by $\xi$).

\begin{figure}[htbp]
    \includegraphics[width=0.7\linewidth]{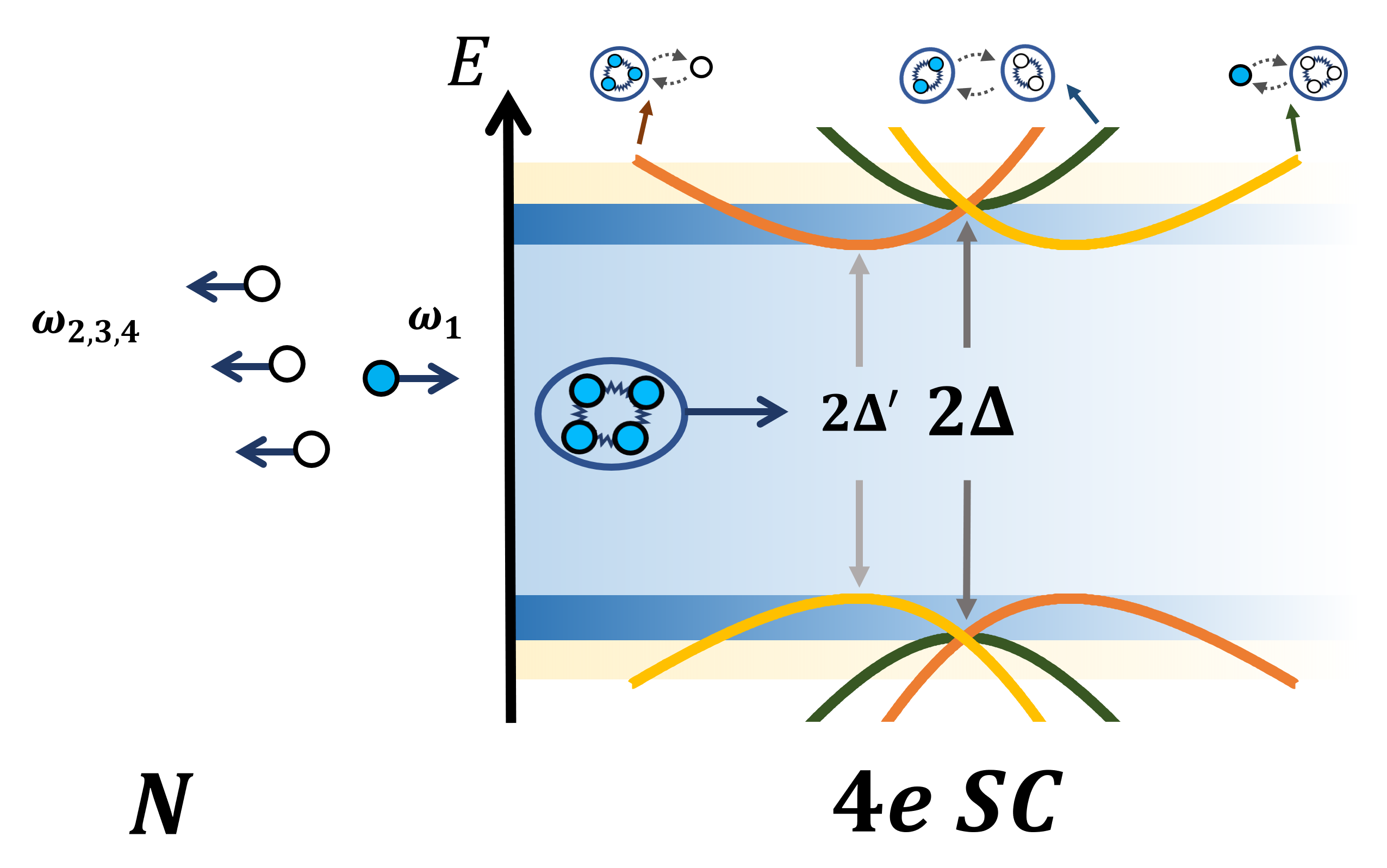}
    \caption{\label{p_4e}
    Schematics of charge-$4e$ Andreev reflection in the normal metal/charge-$4e$ superconductor junction, where one incident electron carrying energy $\omega_1$ can be reflected to three holes carrying energy $\omega_{2,3,4}$ at the interface of the junction, injecting a four-electron pair into the right side.
    The green, yellow and orange lines show the energy dispersion of different types of quasiparticles in charge-$4e$ SC\@.
    While the mixing of charge-$2e$ particles (holes) owns a direct gap $\Delta$, the mixing of charge-$e/3e$ particles (holes) owns an indirect gap $\Delta'=\sqrt{3}\Delta/2$.
    }
\end{figure}

Equiped with this relation, we can write the EOM for charge-$4e$ SC,
\begin{eqnarray}\label{eq_eq}
    (\omega^-_z\ \mathbf{I}_{3} - \mathcal{H})
    \mathbf{F}
    &&= -i \mathbf{F}'_1 ,
\nonumber\\
    (\omega^-_{yz}\ \mathbf{I}_{9} - \mathcal{H}_1)
    \mathbf{F}_1
    &&= -i \mathbf{F}'_2 ,
\nonumber\\
    (\omega^-_{xyz}\ \mathbf{I}_{27} - \mathcal{H}_2)
    \mathbf{F}_2
    &&= -i \mathbf{F}_0 ,
\end{eqnarray}
with $\omega^-_z=\omega_z-i\eta, \omega^-_{yz}=\omega_y+\omega_z-i\eta, \omega^-_{xyz}=\omega_x+\omega_y+\omega_z-i\eta$ and $\eta$ measuring the energy relaxation rate\cite{cuevas_Hamiltonian_1996}.
$\mathbf{I}_n$ is the $n\times n$ identity matrix.
$\mathcal{H}, \mathcal{H}_{1,2}$ are the coefficient matrix and $\mathbf{F}, \mathbf{F}_{0,1,2}$ are the vectors containing charge-$4e$ anomalous Green's functions with different number of time variables.
The superscript $'$ means to select the first $3(9)$ elements of $\mathbf{F}_1(\mathbf{F}_2)$.
We leave the explicit form of $\mathcal{H}, \mathcal{H}_{1,2}$ and $\mathbf{F}, \mathbf{F}_{0,1,2}$ in Appendix~\ref{eom_eq} and show here the result of $\mathscr{F}$ at zero temperature,
\begin{widetext}
\begin{eqnarray}
    \mathscr{F}(\omega_x, \omega_y, \omega_z)
    = i\Delta \frac{
        (\omega^-_z + 3\epsilon_k)
        (\omega^-_{yz} + 2\epsilon_k)
        (\omega^-_{xyz} + \epsilon_k)
        + \Delta^2
        ( \omega^-_{z} +\omega^-_{yz} +\omega^-_{xyz} +2\epsilon_k )
    }{
        \Big[
            {(\omega^-_{z}+\epsilon_k)}^2 - E^2_k
        \Big]
        \Big[
            \omega^{-2}_{yz} - E^2_k
        \Big]
        \Big[
            {(\omega^-_{xyz} -\epsilon_k)}^2-E^2_k
        \Big]
        (\omega^-_{xyz} + \epsilon_k)
    } ,
\end{eqnarray}
\end{widetext}
with $E_k^2=4\epsilon_k^2+\Delta^2$.
This gives us more insights into the properties of charge-$4e$ SC, especially the energy dispersion of quasiparticles, which can be obtained from poles of $\mathscr{F}$.
Similar to the charge-$2e$ SC, the energy dispersion here also shows a gapped feature.
However, we note that there are two types of gaps.
As shown in Fig.\ref{p_4e}, while $\pm E_k$ (shown as green lines) contribute a direct gap of $\Delta$ for the mixing of charge-$2e$ particles (holes), $\pm\epsilon_k\pm E_k$ (shown as yellow and orange lines) contribute an indirect gap of $\Delta'=\sqrt{3}\Delta/2$ for the mixing of charge-$e/3e$ particles (holes).
Similar to the charge-$2e$ SC where the Andreev reflection can be enhanced at the gap, we will show below that the peaks of Andreev reflection in charge-$4e$ SC are also related to those gaps.

We then consider adding the coupling of left leads into $\mathscr{F}$.
In general, the introducing of lead coupling would mix the momentum in the right side and make $k$ not a good quantum number anymore.
Besides, the nonzero voltage bias and the flowing current could regulate the behavior of quasiparticles through the occupation numbers [see in Eq. (\ref{eq_com}) and Appendix~\ref{eom_neq}].
These things would bring a lot of complexity to the description of the transport in charge-$4e$ SC\@.
Therefore, to capture the main characteristics of charge-$4e$ AR, we consider below an approximation with weak coupling, which can also be achieved in experiments conveniently.
At this point, the momentum mixture is weak as well as the current is relatively small so that we can see the lead coupling as an extra self-energy $\mathcal{E}$ to the charge-$4e$ anomalous Green's function while keep other things unchanged.
This would make a correction to the EOM in Eq. (\ref{eq_eq})
\begin{eqnarray}\label{eq_neq}
    (\omega^-_z\ \mathbf{I}_{3} - \mathcal{H} - \mathcal{E})
    \mathbf{F}_{neq}
    &&= -i \mathbf{F}'_{1,neq} ,
\nonumber\\
    (\omega^-_{yz}\ \mathbf{I}_{9} - \mathcal{H}_1 - \mathcal{E}_1)
    \mathbf{F}_{1,neq}
    &&= -i \mathbf{F}'_{2,neq} ,
\nonumber\\
    (\omega^-_{xyz}\ \mathbf{I}_{27} - \mathcal{H}_2  - \mathcal{E}_2)
    \mathbf{F}_{2,neq}
    &&= -i \mathbf{F}_0 ,
\end{eqnarray}
with the subscript `neq' distinguishing the results here from that in equilibrium case.
We leave the explicit form of $\mathcal{E}, \mathcal{E}_{1,2}$ and discussion in Appendix~\ref{eom_neq}.

\section{\label{ar}charge-4e Andreev reflection}

\begin{figure}[htbp]
    \includegraphics[width=1.0\linewidth]{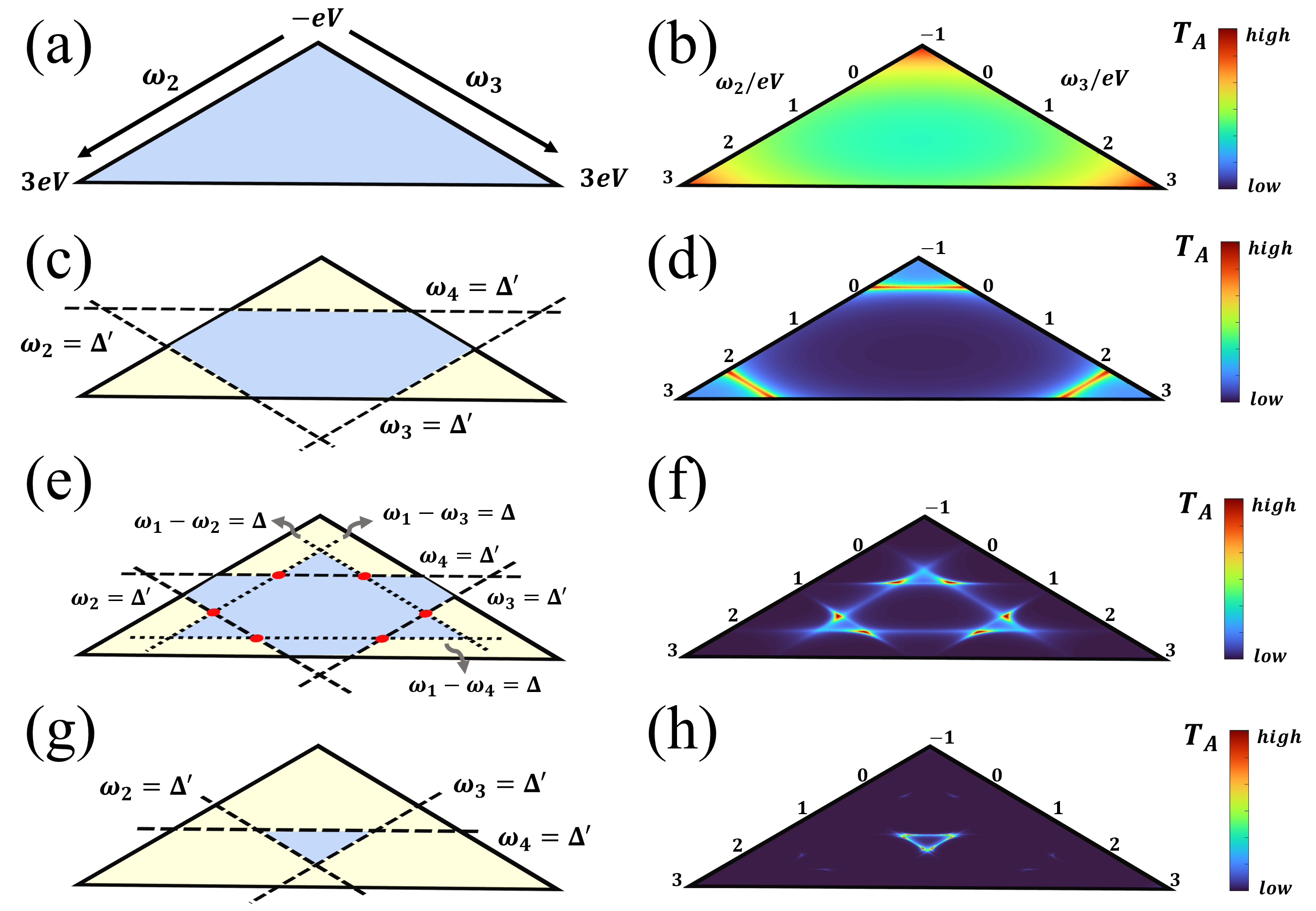}
    \caption{\label{p_ta}
    Schematics of charge-$4e$ Andreev reflection in $\omega_2-\omega_3$ energy plane (left column) with corresponding numerical results of $T_A$ at $eV=0.2, 0.4, 0.8, 1.8$ (from top to bottom of the right column) in the weak coupling case.
    While the blue regions denote the ``in-gap'' areas, we distinguish the separation lines induced by different gaps with different types of dotted lines.
    Moreover, we set $\Delta=1, \rho=1, \Gamma^L=0.02$ and $\eta=0.01$ in calculations\cite{kaplan_Quasiparticle_1976}.}
\end{figure}

We now reach the stage to give the whole picture of the charge-$4e$ AR, together with the calculated results of Andreev coefficient $T_{A}$.
Unlike the charge-$2e$ AR where the energy of the reflected hole is constrained to that of the incident electron,
the loose constraint in charge-$4e$ AR would give two additional energy degrees of freedom $\omega_2,\omega_3$ if we consider the incident electron with certain energy $\omega_1$, leaving $\omega_4=\omega_1-\omega_2-\omega_3$.
Together with Eq. (\ref{eq_ga}) which shows that the contribution of $T_{A}$ to $G_{A}$ is further constrained by Fermi distribution functions,
for a given $\omega_1=eV>0$, we indeed only need to investigate the behavior of $T_{A}$ in $-eV<\omega_{2,3,4}<3eV$, which forms a triangle region in the $\omega_2-\omega_3$ plane [see in Fig.\ref{p_ta} (a)].
Besides, the two types of gaps shown in Sec.\ref{nonper} would bring different ``in-gap'' areas to the energy triangle.
We find that similar to the charge-$2e$ AR, the charge-$4e$ AR would also be enhanced around the energy gap as well as decay quickly outside the gap.
Therefore, as $\omega_1$ varies, the competing of these two gaps would change the ``in-gap'' area, which dominants the behavior of $T_{A}$ and eventually influences the result of conductance $G_{A}$.

We begin with $\omega_1$ being relatively small.
As shown in Fig.\ref{p_ta} (a), the whole energy triangle is inside the gap $\Delta'$ with all the incident (reflected) particles (holes) lying below the gap.
Therefore, $T_A$ stays stable with different energy but is slightly smaller in the center than around [see in Fig.\ref{p_ta} (b)].
If we improve the strength of coupling, $T_A$ would increase in the whole area.
As $\omega_1$ increases, some incident (reflected) particle (hole) would reach the gap $\Delta'$ where Andreev reflection gets enhanced.
This requires that the maximum energy allowed to single particle (hole) should satisfy $3eV\ge\Delta'$, giving a separation point $\omega_{c1}=\Delta'/3$.
When $\omega_1>\omega_{c1}$, the whole area would be separated by three lines $\omega_{2,3,4}=\Delta'$ located at the gap of charge-$e/3e$ particles [see in Fig.\ref{p_ta} (c)].
At weak coupling case, the strength of $T_A$ concentrates on these gap lines [see in Fig.\ref{p_ta} (d)].
As we improve the strength of coupling, those $T_A$ belongs to the ``in-gap'' area [shown as the blue region in Fig.\ref{p_ta} (c)] would also increase.

The further increasing of $\omega_1$ allows that two of the  electrons can carry a total energy of $\Delta$, touching the gap of charge-$2e$ particles.
This requires that $2eV\ge\Delta$, leading to another separation point $\omega_{c2}=\Delta/2$.
When $\omega_1$ exceeds $\omega_{c2}$, a new set of three lines appears, with $\omega_1-\omega_2=\Delta , \omega_1 - \omega_3=\Delta$ and $\omega_1 - \omega_4 = \Delta$ located at the gap of charge-$2e$ particles.
At this time, the ``in-gap'' area is encircled by six lines, showing a Star of David structure [see in Fig.\ref{p_ta} (e)].
At weak coupling case, the peaks of $T_A$ mainly focuses on the crossover points of these lines [see in Fig.\ref{p_ta} (f)].
The improvement of coupling would enhance the $T_A$ in the whole area inside the Star of David structure.

As $\omega_1$ further increases across $\omega_{c3}=2\Delta-\Delta'$, the
``in-gap'' area brought by $\Delta$ fully enters into the area brought by $\Delta'$.
As shown in Fig.\ref{p_ta} (g) and (h), the maximum of $T_A$ is then back to the position of $\Delta'$ gap.
This ``in-gap'' area would eventually disappear as $\omega_1\ge 3\Delta'$, leaving the maximum of $T_A$ at the center of the energy triangle $\omega_2=\omega_3=\omega_4=\omega_1/3$.
This also matches the Eq. (\ref{eq_wd}) from the perturbation
theory in the case of $\Delta\ll\omega_1$, showing the consistency of results provided by different methods.

\section{\label{cond}Conductance}

\begin{figure}[htbp]
    \includegraphics[width=0.95\linewidth]{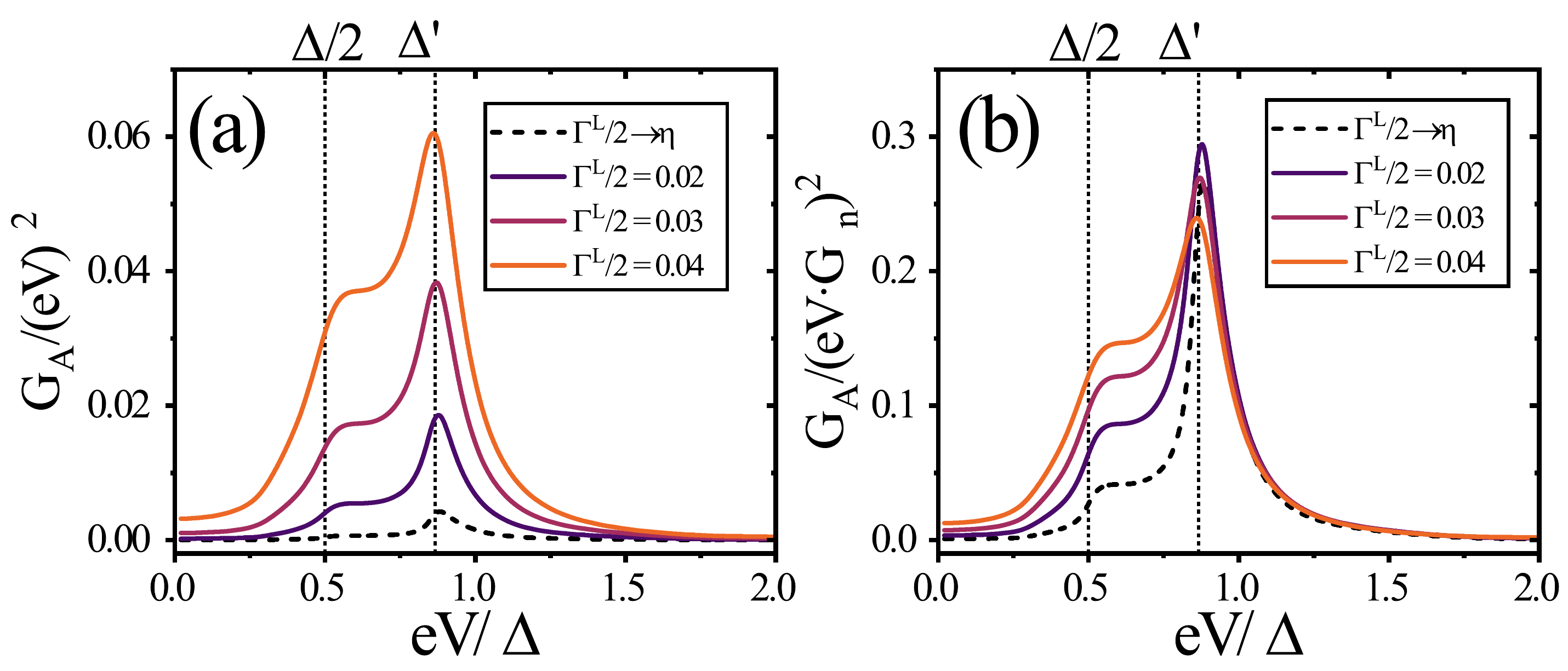}
    \caption{\label{p_ga}
    (a) Conductance contributed by the charge-$4e$ AR with different coupling strength.
    (b) shows the consistency of the results of the weak coupling with that of the weak coupling limit, where the nonequilibrium Green's functions are replaced by the equilibrium Green's functions.
    We use dotted lines to highlight the major features in $G_A$.
    Besides, $G_n\approx4\pi^2\rho^2|t_c|^2= 2\pi\rho\Gamma^L$ is the normal state conductance in weak coupling case\cite{blonder_Transition_1982}.
    We omit the unit $e^2/h$ for conductance here and below, and keep other parameters the same as those in Fig.\ref{p_ta} in calculations.}
\end{figure}

Clarifying the behavior of $T_A$ helps us in understanding the conductance of charge-$4e$ SC, which can be measured in experiments.
We first consider the conductance $G_A$ contributed by the charge-$4e$ AR [see in Fig.\ref{p_ga} (a)].
From the numerical results, we find that the $G_A$ possesses two major features.
One is the conductance peak located at $\Delta'=\sqrt{3}\Delta/2$.
The other is the emerging plateau beginning from $\Delta/2$.

\begin{figure*}[htbp]
    \includegraphics[width=0.95\linewidth]{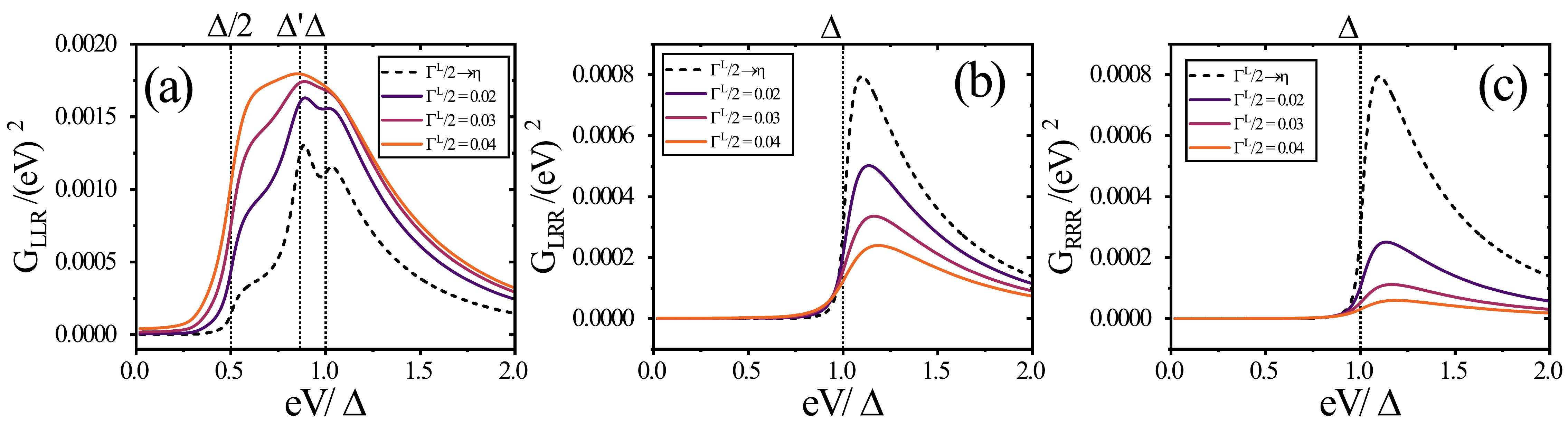}
    \caption{\label{p_gabc}
    Conductance contributed by the charge-$4e$ AR with some reflected holes entering the right leads.
    (a), (b) and (c) show the conductance $G_{LLR}, G_{LRR}$ and $G_{RRR}$, respectively, with different coupling strength.
    We use dotted lines to highlight major features in those conductances and keep other parameters the same as those in Fig.\ref{p_ta} in calculations.}
\end{figure*}

When we consider increasing $eV$ from zero, at first, $G_A$ is contributed by those ``in-gap'' processes [shown in Fig.\ref{p_ta} (a) and (b)].
At weak coupling case, they are relatively small, resulting in a near zero $G_A$.
Once $eV$ surpasses $\omega_{c1}$, the emerging gap lines enhance the Andreev reflection [shown in Fig.\ref{p_ta} (c) and (d)], causing the increasing of $G_A$ before $eV=\Delta/2$.
The further increasing of $eV$, however, turns the maximum of $T_A$ from lines to points.
The emerging plateau is then attributed to the fixed number of $T_A$ peaks [shown in Fig.\ref{p_ta} (e) and (f)].
Such peaks would reach their maximum at $eV=\Delta'$, where all the participating electrons touch the gaps $\pm\Delta'$.
We therefore get the $G_A$ peak located at $eV=\Delta'$, which differs from the charge-$2e$ AR that $G_A$ is enhanced at $\Delta$\cite{blonder_Transition_1982}.
Apart from the peak, $G_A$ decays.
While in charge-$2e$ AR, a quick $\Delta^2/[\Delta^2-{(eV)}^2]$ behavior is presented\cite{blonder_Transition_1982},
the asymmetric behavior here between $eV=\Delta'$ (especially when $\Gamma^L/2\rightarrow\eta$) shows the slow dacaying in charge-$4e$ $G_A$ due to the integral nature of it.
Moreover, we compare our results with decreasing the coupling $\Gamma^L$, finding that the peak and plateau always appear at weak coupling case.
As the coupling desreases, the nonequilibrium Green's functions would regress to equilibrium Green's functions.
Our results in Fig.\ref{p_ga} (b) then indicate good consistency, with the small values of $G_A$ being scaled by $G^2_n\propto{(\Gamma^L)}^2$ at weak coupling case.

We then consider the contributions of other processes to the total conductance [see Appendix~\ref{tran_4e} for the formulas of the 
other processes].
Except the quasiparticle tunneling process, the other processes can also be viewed as Andreev reflection with some of the reflected holes entering into the right leads.
Similar to the charge-$2e$ AR where an incident electron can be converted into a hole in the superconductor due to the ``branch crossing''\cite{blonder_Transition_1982,cuevas_Hamiltonian_1996},
we distinguish those processes by which side the reflected holes enter and use $T_{\alpha\beta\gamma}, G_{\alpha\beta\gamma}$ with $\alpha,\beta,\gamma=L,R$ to denote them.
Due to the coexistence of left leaving holes and right leaving holes, we find that they possesses features both in Andreev reflection and normal tunneling with different types of quasiparticles.
Therefore, we cannot simply attributed them to the transmission processes like what we usually do in charge-$2e$ SC\cite{sun_Quantum_2009,lv_Magnetoanisotropic_2018}.

Fig.\ref{p_gabc} (a), (b), and (c) show numerical results for
the conductance $G_{LLR}$, $G_{LRR}$, and $G_{RRR}$, respectively. 
These processes contain both features in Andreev reflection and quasiparticle tunneling.
While the emerging increasing from $eV=\Delta/2$ and the peak at $eV=\Delta'$ is similar to that behavior of $G_A$, we attribute the peak at $eV=\Delta$ to the combination of charge-$2e$ quasiparticles in the right leads.
Note that the symmetry preserves that $G_{LLR}=G_{LRL}=G_{RLL}$ and $G_{LRR}=G_{RLR}=G_{RRL}$.

\begin{figure}[htbp]
    \includegraphics[width=0.8\linewidth]{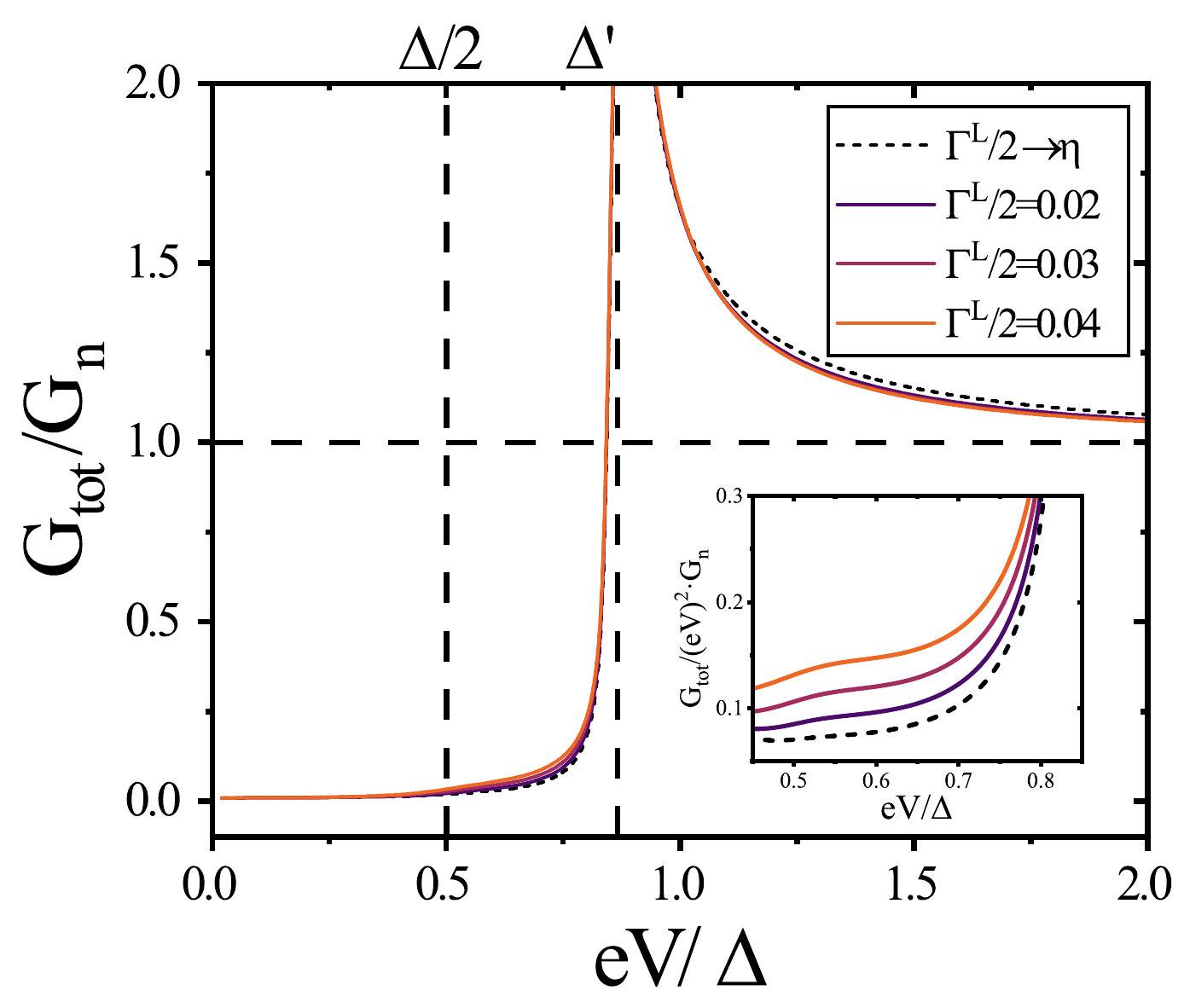}
    \caption{\label{p_gtot}
    Total conductance with different coupling strength.
    We use dashed lines to highlight the major features in $G_{tot}$ and the insert shows detailed behavior of the plateau.
    Besides, we keep other parameters the same as those in Fig.\ref{p_ta} in calculations.}
\end{figure}

Finally, combining all the processes (including all Andreev reflections $G_{\alpha\beta\gamma}$ and the quasiparticle tunneling $G_q$) gives us the total differential conductance $G_{tot}$.
As shown in Fig.\ref{p_gtot}, at weak coupling case, we have an U-shape curve of $G_{tot}$.
We also find the plateau beginning from $eV=\Delta/2$ as well as the conductance peak at $eV=\Delta'$.
While the former is mainly contributed by $G_A$ and $G_{LLR,LRL,RLL}$,
the latter is mainly contributed by the divergence of the DOS in charge-$4e$ SC\cite{li_Charge_2022}.
As $eV$ increases away from the gap, $G_{tot}$ eventually tends to the normal state conductance $G_n$, backing to the results in normal metal/normal metal junctions\cite{cuevas_Hamiltonian_1996}.

Moreover, it is beneficial to discuss the possible behaviors of $G_{\alpha\beta\gamma}$ in strong coupling case.
When we improve the coupling, a direct result is the increasing of $T_A$ in those ``in-gap'' area, which would suppress the contribution in gap lines.
Therefore, when $eV$ is close to zero, $G_A/{(eV)}^2\approx(2/\pi^2)T_A$ the constant proportional to the area of the energy triangle.
As $eV$ increases, the behavior of $G_A/{(eV)}^2$ would be proportional to the area of the ``in-gap'' region, and the total curve of $G_A$ would show a V-shape behavior.
Besides, to see the possible values of those conductances, we further considering an incident electron with energy $E$, the conservation of probability in charge-$4e$ SC case requires that
\begin{eqnarray}
    1 = T_q(E) + R(E)
    + \sum_{\alpha\beta\gamma} \int'_{1234-i}
    T_{\alpha\beta\gamma} \big|_{\omega_i^e=E} ,
\end{eqnarray}
where $T_q$ is the tunneling coefficient and 
$R$ is the reflection coefficient.
$T_{\alpha\beta\gamma}$ are Andreev reflection coefficients with $T_{LLL}=T_A$.
Due to the non-negativity of those coefficients, we have
\begin{eqnarray}\label{eq_cons}
    0\le\ \int'_{1234-i}
    T_{\alpha\beta\gamma} \big|_{\omega_i^e=E}\ \le 1 ,
\end{eqnarray}
which would give us an estimation for $G_{\alpha\beta\gamma}$.
Note that here we donot involve Fermi distribution functions.
This would extend the integral range to the whole energy plane and cause the result of single integral in $G_{\alpha\beta\gamma}$ [see in Eq. (\ref{eq_ga}) and (\ref{eq_gabc})] less than that in Eq. (\ref{eq_cons}).
Therefore, we have
\begin{eqnarray}
    G_{\alpha\beta\gamma}
    \le (n_{\alpha\beta\gamma}+1)\frac{e^2}{h},
\end{eqnarray}
with $n_{\alpha\beta\gamma}$ measuring the number of holes entering the left leads.
We then expect to approach these values at some strong coupling cases.

\section{\label{diss}discussion and conclusions}

The calculations above provide a comprehensive description for the charge-$4e$ AR, from the transport formula to the picture of the behavior of $T_A$.
Differing from the charge-$2e$ SC, in charge-$4e$ SC,  the superconducting potential $\Delta$ introduces two types of gaps, an indirect gap of $\Delta'=\sqrt{3}\Delta/2$ for charge-$e/3e$ quasiparticles and a direct gap of $\Delta$ for charge-$2e$ quasiparticles.
This two gaps compete with each other to decide the ``in-gap'' area for the $T_A$, thus introducing the conductance peak at $eV=\Delta'$ and the plateau beginning from $eV=\Delta/2$ at weak coupling case.

In experiments, these features can help us distinguish charge-$4e$ SC from the charge-$2e$ SC\@.
As the value of the supercondcutoring potential usually remains unknown, it is hard to distinguish $\Delta'$ from $\Delta$ directly.
Therefore, the existence of the plateau not only gives us a direct evidence that differs from the conductance spectrum of charge-$2e$ SC but also provide us a way to check the behavior of $\sqrt{3}$ as the ratio of the position of the gap and plateau.
Moreover, if we can change the coupling continously, like approaching the tip to the sample in STM experiments, the rising of the conductance inside the gap can alter the curve of the differential conductance from U-shape and plateau to V-shape, with a parabolic behavior near the center of the gap.
This would give more hallmarks for experimental observation.

In summary, we investigate the Andreev reflection in a normal metal/charge-$4e$ superconductor junction, which involves four particles due to the characteristics of quartet condensation.
Using nonequilibrium Green's function method, we obtain a four-particle-type Laudauer-B\"uttiker formula with generalized charge-$4e$ anomalous Green's functions to describe this process.
We then calculate and clarify the behavior of this process with various incident energy and show the conductance contributed by it.
Our results can enrich the understanding of the transport with charge-$4e$ SC and give more guidance for the future experimental verifications.

\begin{acknowledgments}

This work was financially supported by the National Natural Science Foundation of China (Grant No. 12374034 and No. 11921005),
the Innovation Program for Quantum Science and Technology (2021ZD0302403),
and the Strategic priority Research Program of Chinese Academy of Sciences (Grant No.XDB28000000).
We also acknowledge the Highperformance Computing Platform of Peking University for providing computational resources.

\end{acknowledgments}

\appendix

\section{Derivation of the Formula for Transport Processes in Normal Metal/Supercondcutor Junctions}

In this Appendix, we derive the formula for transport processes in the normal metal/supercondcutor junction, with the superconducting side being either charge-$2e$ supercondcutor or charge-$4e$ supercondcutor.

\subsection{\label{tran_2e}Transport in Charge-2e Superconductivity}

We first derive the transport formula in the case of charge-$2e$ SC\@.
In order to describe the Andreev reflection, we explicitly insert the coupling term of channel 1$\downarrow$ into $G_{1\uparrow,k_1k_1'}(\tau_1,\tau_1')$.
By using the Wick's theorem, we get
\begin{widetext}
\begin{equation}
    \big[
        \mbox{part of }
        G_{1\uparrow,k_1k_1'}(\tau_1,\tau_1')
    \big]
    = \int_c d\tau_2 d\tau_2'
    \sum_{k_2k_2'}
    F_{k_1k_2}(\tau_1,\tau_2)\ 
    \Sigma_{1\downarrow}(\tau_2,\tau_2')\ 
    F^\dagger_{k_2'k_1'}(\tau_2',\tau_1') ,
\end{equation}
with $F_{k_1k_2}(\tau_1,\tau_2)$ and $F^\dagger_{k_2'k_1'}(\tau_2',\tau_1')$ being the nonequilibrium anomalous Green's functions.
$\Sigma_{1\downarrow}(\tau_2,\tau_2') = \sum_{k} |t_c|^2 g^h_{1\downarrow,k}(\tau_2,\tau_2')$ is the self-energy of coupling to channel 1$\downarrow$
with $g^h_{1\downarrow,k}(\tau_2,\tau_2') = -i\langle a^\dagger_{1-k\downarrow}(\tau_2) a_{1-k\downarrow}(\tau_2') \rangle^c_0 $ being the hole-type free Green's function in the decoupled system (i.e., when $t_c=0$).

We then apply the analytic continuation\cite{haug_Quantum_2008} to get the less (greater) Green's function $G^{\lessgtr}_{1\uparrow,kk'}(\omega_1)$ as the Fourier transform of $G^{\lessgtr}_{1\uparrow,k_1k_1'}(t,0)$
\begin{eqnarray}\label{eq_gg2e}
    &&\big[
        \mbox{part of }
        G^<_{1\uparrow,k_1k_1'}(\omega_1)
    \big]
    = \sum_{k_2k_2'} \sum_{ss'} {(-1)}^P
    F^{+s}_{k_1k_2}(\omega_1)
    \
    \Sigma^{ss'}_{1\downarrow}(\omega_1)
    \
    F^{\dagger,s'-}_{k_2'k_1'}(\omega_1) ,
    \nonumber\\
    &&\big[
        \mbox{part of }
        G^>_{1\uparrow,k_1k_1'}(\omega_1)
    \big]
    = \sum_{k_2k_2'} \sum_{ss'} {(-1)}^P
    F^{-s}_{k_1k_2}(\omega_1)
    \
    \Sigma^{ss'}_{1\downarrow}(\omega_1)
    \
    F^{\dagger,s'+}_{k_2'k_1'}(\omega_1) ,
\end{eqnarray}
with $s,s'=\pm$ being the branch index and $P$ being the total number of `$-$' branch among them.
Besides, $F^{\pm\pm}_{k_1k_2}(\omega_1), F^{\dagger,\pm\pm}_{k_2'k_1'}(\omega_1)$ and $\Sigma^{\pm\pm}_{1\downarrow}(\omega_1)$ are Fourier transforms of $F_{k_1k_2}(t^\pm,0^\pm), F^{\dagger}_{k_2'k_1'}(t^\pm,0^\pm)$ and $\Sigma_{1\downarrow}(t^\pm,0^\pm)$, respectively.
In order to get $G^{\lessgtr}_{1\uparrow,A}$,
we need to select the less (greater) part $\Sigma^{\lessgtr}_{1\downarrow}$
from $\Sigma^{\pm\pm}_{1\downarrow}$.
By using $\Sigma^{r}_{1\downarrow} =-\Sigma^{a}_{1\downarrow} =-i\Gamma^L/2$,
$\Sigma^{<}_{1\downarrow} =if^h_{i\downarrow}\Gamma^L$, and
$\Sigma^{>}_{1\downarrow} =-i\bar{f}^h_{i\downarrow}\Gamma^L$\cite{rammer_Quantum_1986}
we eventually obtain Eq. (\ref{eq_g2e}) and (\ref{eq_fr2e}) from Eq. (\ref{eq_gg2e}).

While Eq. (\ref{eq_gg2e}) contains the process of Andreev reflection, it should be emphasized that the rest parts of $\Sigma^{\pm\pm}_{1\downarrow}$ (i.e., $\Sigma^{r/a}_{1\downarrow}$) in Eq. (\ref{eq_gg2e}) also contributes to the other transport processes.
Using the Keldysh equation\cite{cuevas_Hamiltonian_1996}
\begin{eqnarray}
    \mathbf{G}^{\lessgtr}
    &&=(1+\mathbf{G}^r\mathbf{\Sigma}^r)
    \mathbf{G}_0^{\lessgtr}
    (1+\mathbf{\Sigma}^a\mathbf{G}^a)
    +
    \mathbf{G}^r
    \mathbf{\Sigma}^\lessgtr
    \mathbf{G}^a
    =\mathbf{G}^r
    \left[
        \mathbf{G}_0^{r-1}
        \mathbf{G}_0^\lessgtr
        \mathbf{G}_0^{a-1}
        +
        \mathbf{\Sigma}^\lessgtr
    \right]
    \mathbf{G}^a ,
\end{eqnarray}
the exact form of $G^{\lessgtr}_{1\uparrow,k_1k_1'}$ can be derived as\cite{sun_Resonant_1999,sun_Control_2000,wang_Statistical_2003} (omitting the energy variable $\omega_1$)
\begin{eqnarray}\label{eq_glg}
    &&G^{<}_{1\uparrow,k_1k_1'}
    = \sum_{k_2k_2'}
    G^{r}_{1\uparrow,k_1k_2}
    \left[
        if^e_{1\uparrow} \Gamma^L
        + if \Gamma^R \delta_{k_2k_2'}
    \right]
    G^{a}_{1\uparrow,k_2'k_1'}
    + F^{r}_{k_1k_2}
    \left[
        if^h_{1\downarrow} \Gamma^L
        + if \Gamma^R \delta_{k_2k_2'}
    \right]
    F^{\dagger,a}_{k_2'k_1'} ,
    \nonumber\\
    &&G^{>}_{1\uparrow,k_1k_1'}
    = \sum_{k_2k_2'}
    G^{r}_{1\uparrow,k_1k_2}
    \left[
        -i\bar{f}^e_{1\uparrow} \Gamma^L
        -i\bar{f} \Gamma^R \delta_{k_2k_2'}
    \right]
    G^{a}_{1\uparrow,k_2'k_1'}
    + F^{r}_{k_1k_2}
    \left[
        -i\bar{f}^h_{1\downarrow} \Gamma^L
        -i\bar{f} \Gamma^R \delta_{k_2k_2'}
    \right]
    F^{\dagger,a}_{k_2'k_1'} ,
\end{eqnarray}
\end{widetext}
where $\bar{f}=1-f$ and $f=f(\omega_1)$ is the Fermi distribution function in the right lead.
$\Gamma^R=2\eta$ measures the linewidth of the quasiparticles in the right lead\cite{cuevas_Hamiltonian_1996}.
Substituted Eq. (\ref{eq_glg}) into Eq. (\ref{eq_current}), we then get the transport formula for other processes.

One is the quasiparticle tunneling process, contributing a current
\begin{equation}\label{eq_jq}
    I_{1\uparrow,q}
    = \frac{e}{h} \int \mathrm{d}\omega_1
    (f^e_{1\uparrow} - f)\ T_q ,
\end{equation}
where $T_q(\omega_1) = \Gamma^L \Gamma^R \sum_{k_1k_2k_1'} G^r_{1\uparrow,k_1k_2} G^a_{1\uparrow,k_2k_1'}$ is the tunneling coefficient.

The other is the Andreev process with the reflected hole entering the right lead, contributing a current
\begin{eqnarray}\label{eq_jr}
    I_{1\uparrow,R}&
    = &\frac{e}{h} \int \mathrm{d}\omega_1
    (f^e_{1\uparrow}\bar{f} - \bar{f}^e_{1\uparrow} f)\ T_R \nonumber\\
    &= & \frac{e}{h} \int \mathrm{d}\omega_1
    (f^e_{1\uparrow} - f)\ T_R ,
\end{eqnarray}
with $T_R(\omega_1)=\Gamma^L\Gamma^R \sum_{k_1k_2k_1'} F^r_{k_1k_2} F^{\dagger,a}_{k_2k_1'}$.

In fact, we usually donot distinguish these two processes and define the transmission current\cite{sun_Quantum_2009}
\begin{equation}
    I_{1\uparrow,trans}
    = \frac{e}{h} \int \mathrm{d}\omega_1
    (f^e_{1\uparrow} - f)\ T_{trans} ,
\end{equation}
with transmission coefficient $T_{trans}=T_q+T_R$.
Together with Eq. (\ref{eq_ja_2e}), we have the total current
\begin{equation}
    I_{1\uparrow,tot} = I_{1\uparrow,A} + I_{1\uparrow,trans} .
\end{equation}

\subsection{\label{tran_4e}Transport in Charge-4e Supercondcutivity}

Similar to the charge-$2e$ AR, the charge-$4e$ AR can be described by inserting the coupling terms of other three channels into $G_{1\uparrow,k_1k_1'}(\tau_1,\tau_1')$.
Using the Wick's theorem, we get
\begin{widetext}
\begin{eqnarray}
    &&\big[
        \mbox{part of }
        G_{1\uparrow,k_1k_1'}(\tau_1,\tau_1')
    \big]
    = -\int_c
    d\tau_2 d\tau_3 d\tau_4
    d\tau_2' d\tau_3' d\tau_4'
    \sum_{\substack{k_2k_3k_4\\k_2'k_3'k_4'}}
    \nonumber\\
    &&\qquad\qquad
    F_{k_1k_2k_3k_4}
    (\tau_1,\tau_2,\tau_3,\tau_4)
    \
    \Sigma_{1\downarrow} (\tau_2,\tau_2')
    \Sigma_{2\uparrow} (\tau_3,\tau_3')
    \Sigma_{2\downarrow} (\tau_4,\tau_4')
    \
    F^\dagger_{k_4'k_3'k_2k_1'}
    (\tau_4',\tau_3',\tau_2',\tau_1') ,
\end{eqnarray}
with $F_{k_1k_2k_3k_4} (\tau_1,\tau_2,\tau_3,\tau_4)$ and $F^\dagger_{k_4'k_3'k_2'k_1'}(\tau_4',\tau_3',\tau_2',\tau_1')$ being the nonequilibrium charge-$4e$ anomalous Green's functions.
$\Sigma_{i\sigma} (\tau,\tau') = \sum_{k} |t_c|^2 g^h_{i\sigma,k} (\tau,\tau')$ is self-energy of coupling to channel $i\sigma$
with $g^h_{i\sigma,k}(\tau,\tau') = -i\langle a^\dagger_{ik\sigma}(\tau) a_{ik\sigma}(\tau') \rangle^c_0 $ being the hole-type free Green's function in the decoupled system (i.e., when $t_c=0$).

We then apply the analytic continuation\cite{haug_Quantum_2008} and use the Fourier transform to get the less (greater) Green's functions
\begin{eqnarray}\label{eq_gg4e}
    &&\big[
        \mbox{part of }
        G^<_{1\uparrow,k_1k_1'}(\omega_1)
    \big]
    = -\int_{234}
    \sum_{\substack{k_2k_3k_4\\k_2'k_3'k_4'}}\
    \sum_{\substack{s_2s_3s_4\\s_2's_3's_4'}}
    {(-1)}^P
    F^{+s_2s_3s_4}_{k_1k_2k_3k_4}
    \Sigma^{s_2s_2'}_{1\downarrow}(\omega_2)
    \Sigma^{s_3s_3'}_{2\uparrow}(\omega_3)
    \Sigma^{s_4s_4'}_{2\downarrow}(\omega_4)
    F^{\dagger,s_4's_3's_2'-}_{k_4'k_3'k_2'k_1'},
    \nonumber\\
    &&\big[
        \mbox{part of }
        G^>_{1\uparrow,k_1k_1'}(\omega_1)
    \big]
    = -\int_{234}
    \sum_{\substack{k_2k_3k_4\\k_2'k_3'k_4'}}\
    \sum_{\substack{s_2s_3s_4\\s_2's_3's_4'}}
    {(-1)}^P
    F^{-s_2s_3s_4}_{k_1k_2k_3k_4}
    \Sigma^{s_2s_2'}_{1\downarrow}(\omega_2)
    \Sigma^{s_3s_3'}_{2\uparrow}(\omega_3)
    \Sigma^{s_4s_4'}_{2\downarrow}(\omega_4)
    F^{\dagger,s_4's_3's_2'+}_{k_4'k_3'k_2'k_1'},
\end{eqnarray}
\end{widetext}
with $s_{2,3,4},s_{2,3,4}'=\pm$ being the branch index and $P$ denoting the total number of `$-$' branch among them.
$F^{\pm\pm\pm\pm}_{k_1k_2k_3k_4} = F^{\pm\pm\pm\pm}_{k_1k_2k_3k_4}(\omega_2,\omega_3,\omega_4)$ and $F^{\dagger,\pm\pm\pm\pm}_{k_4'k_3'k_2'k_1'}=F^{\dagger,\pm\pm\pm\pm}_{k_4'k_3'k_2'k_1'}(\omega_4,\omega_3,\omega_2)$ are the Fourier transforms of $F_{k_1k_2k_3k_4}(0^\pm,t_2^\pm,t_3^\pm,t_4^\pm)$ and $F^{\dagger}_{k_4'k_3'k_2'k_1'}(t_4^\pm,t_3^\pm,t_2^\pm,0^\pm)$.
$\Sigma^{\pm\pm}_{i\sigma}(\omega_j)$ is the Fourier transform of $\Sigma_{i\sigma}(t_j^\pm,0^\pm)$.
In order to get $G^{\lessgtr}_{1\uparrow,A}$, we also need to select the less (greater) part $\Sigma^{\lessgtr}_{i\sigma}$ from the $\Sigma^{\pm\pm}_{i\sigma}$.
With some simplification, we can get Eq. (\ref{eq_g4e}) and (\ref{eq_fr4e}) in the end.

Similar to the charge-$2e$ SC, here we have extra processes contributing to the total conductance, including the quasiparticle tunneling and other three types of Andreev processes.

While the current $I_{1\uparrow,q}$ contributed by the quasiparticle tunneling has the same form as Eq. (\ref{eq_jq}), the formula for other processes can be generalized from Eq. (\ref{eq_jr}).
We denote them as $I_{1\uparrow,\alpha\beta\gamma}$, with $\alpha,\beta,\gamma=L,R$ denoting which side the reflected holes enter, and set $I_{1\uparrow,LLL} = I_{1\uparrow,A}$.
According to the number of reflected holes entering the right side, we have
\begin{eqnarray}
    &&I_{1\uparrow,LLR}
    = \frac{e}{h} \int'_{1234}
    (
        f^{e}f^{e}f^{e}f
        -
        \bar{f}^{e}\bar{f}^{e}\bar{f}^{e}\bar{f}
    )\ T_{LLR} ,
    \nonumber\\
    &&I_{1\uparrow,LRR}
    = \frac{e}{h} \int'_{1234}
    (
        f^{e}f^{e}ff
        -
        \bar{f}^{e}\bar{f}^{e}\bar{f}\bar{f}
    )\ T_{LRR} ,
    \nonumber\\
    &&I_{1\uparrow,RRR}
    = \frac{e}{h} \int'_{1234}
    (
        f^{e}fff
        -
        \bar{f}^{e}\bar{f}\bar{f}\bar{f}
    )\ T_{RRR} ,
\end{eqnarray}
with
\begin{eqnarray}
    &&T_{LLR} = {(\Gamma^L)}^3 \Gamma^R
    \sum_{k_i k_j'}
    F^r_{1\uparrow,k_1k_2k_3k_4}
    F^{\dagger,a}_{1\uparrow,k_4k_3'k_2'k_1'} ,
\nonumber\\
    &&T_{LRR} = {(\Gamma^L)}^2 {(\Gamma^R)}^2
    \sum_{k_i k_j'}
    F^r_{1\uparrow,k_1k_2k_3k_4}
    F^{\dagger,a}_{1\uparrow,k_4k_3k_2'k_1'} ,
\nonumber\\
    &&T_{RRR} = \Gamma^L {(\Gamma^R)}^3
    \sum_{k_i k_j'}
    F^r_{1\uparrow,k_1k_2k_3k_4}
    F^{\dagger,a}_{1\uparrow,k_4k_3k_2k_1'} ,
\end{eqnarray}
Others like $I_{1\uparrow,LRL}(T_{LRL})$, $I_{1\uparrow,RLR}(T_{RLR})$, et al.\ are analogous.

Therefore we can measure their contributions to the conductance by $G_{\alpha\beta\gamma}=\partial I_{1\uparrow,\alpha\beta\gamma}/\partial V$
\begin{eqnarray}\label{eq_gabc}
    &&G_{LLR}
    = \frac{e^2}{h} \sum_{i=1}^3\int'_{1234-i}
    (
        f^e f^e f
        +
        \bar{f}^e\bar{f}^e \bar{f}
    )\ T_{LLR} \Big|_{\omega^e_i=eV} ,
\nonumber\\
\nonumber\\
    &&G_{LRR}
    = \frac{e^2}{h} \sum_{i=1}^2\int'_{1234-i}
    (
        f^e f f
        +
        \bar{f}^e \bar{f} \bar{f}
    )\ T_{LRR} \Big|_{\omega^e_i=eV} ,
\nonumber\\
\nonumber\\
    &&G_{RRR}
    = \frac{e^2}{h} \int'_{1234-1}
    ( fff + \bar{f}\bar{f}\bar{f} )
    \ T_{RRR} \Big|_{\omega^e_1=eV} ,
\end{eqnarray}
assuming that all $T_{\alpha\beta\gamma}$ are independent of voltage $V$.
Others are analogous.
We then obtain the total current and conductance
\begin{eqnarray}
    &&I_{1\uparrow,tot} = I_{1\uparrow,q}
    + \sum_{\alpha\beta\gamma}
    I_{1\uparrow,\alpha\beta\gamma} ,
\nonumber\\
    &&G_{tot} = G_{q}
    + \sum_{\alpha\beta\gamma}
    G_{\alpha\beta\gamma} ,
\end{eqnarray}
with $G_{LLL}=G_{A}$.

\section{\label{per_other}Perturbative Expansion for Other Processes in Normal Metal/Charge-4e Supercondcutor Junctions}

In this Appendix, we briefly discuss the perturbative expansion for other processes.

The first one is the quasiparticle tunneling process, which is described by Eq. (\ref{eq_jq}).
To calculate the noequilibrium $G^{r/a}_{1\uparrow,k_1k_2}$, we begin with the Dyson equation\cite{haug_Quantum_2008}
\begin{widetext}
\begin{eqnarray}
    &&G^{r/a}_{1\uparrow,k_1k_2} (\omega_1)
    = \delta_{k_1k_2} g^{r/a}_{1\uparrow,k_1} (\omega_1)
    + \sum_{k_1'k_2'}
    g^{r/a}_{1\uparrow,k_1} (\omega_1)
    \left[
        \delta_{k_1'k_2'} \Sigma^{r/a}_{L}
        +
        \Sigma^{r/a}_{1\uparrow,k_1'k_2'} (\omega_1)
    \right]
    G^{r/a}_{1\uparrow,k_2'k_2} (\omega_1) ,
\end{eqnarray}
with $g_{1\uparrow,k}$ being the free Green's function.
$\Sigma^{r/a}_L = \mp i\Gamma^L/2$ is the self-energy of coupling to channel 1$\uparrow$ and
$\Sigma^{r/a}_{1\uparrow,k_1'k_2'} (\omega_1)$ is the Fourier transform of the irreducible self-energy $\Sigma^{r/a}_{1\uparrow,k_1'k_2'} (t,0)$ containing the superconducting interaction and the coupling.
Expanding $\Sigma_{1\uparrow,k_1'k_2'} (\tau,\tau')$ to the lowest order of $\Delta$ and applying the analytic continuation\cite{haug_Quantum_2008}, we have
\begin{eqnarray}
    &&\Sigma^{(1)}_{1\uparrow,k_1'k_2'} (\tau,\tau')
    = -\Delta^2
    \mathcal{G}_{1\downarrow,k_1'k_2'} (\tau,\tau')
    \mathcal{G}_{2\uparrow,k_1'k_2'} (\tau,\tau')
    \mathcal{G}_{2\downarrow,k_1'k_2'} (\tau,\tau'),
    \nonumber\\
    &&\Sigma^{(1),\lessgtr}_{1\uparrow,k_1'k_2'} (t,t')
    = -\Delta^2
    \mathcal{G}^{\lessgtr}_{1\downarrow,k_1'k_2'} (t,t')
    \mathcal{G}^{\lessgtr}_{2\uparrow,k_1'k_2'} (t,t')
    \mathcal{G}^{\lessgtr}_{2\downarrow,k_1'k_2'} (t,t'),
\end{eqnarray}
with
\begin{eqnarray}
    \mathcal{G}_{i\sigma,k_1'k_2'} (\tau,\tau')
    = \delta_{k_1'k_2'} g^h_{i\sigma,k_1'} (\tau,\tau')
    + \int_c d\tau_1 d\tau_1' \sum_{k}
    g^h_{i\sigma,k_1'} (\tau,\tau_1)
    \Sigma_{i\sigma}(\tau_1,\tau_1')
    \mathcal{G}_{i\sigma,kk_2'} (\tau_1',\tau').
\end{eqnarray}
\end{widetext}
Using\cite{haug_Quantum_2008} (omit the subscript for simplification)
\begin{eqnarray}
    &&\Sigma^{(1),r} (t,0)
    = \theta(t) \left[
        \Sigma^{(1),>} (t,0)
        -
        \Sigma^{(1),<} (t,0)
    \right] ,
\nonumber\\
    &&\Sigma^{(1),a} (t,0)
    = \theta(-t) \left[
        \Sigma^{(1),<} (t,0)
        -
        \Sigma^{(1),>} (t,0)
    \right] ,
\end{eqnarray}
the Fourier transform to $\Sigma^{(1),r/a} (\omega_1)$ would then lead to integrals over term like $(f_\alpha f_\beta f_\gamma + \bar{f}_\alpha \bar{f}_\beta \bar{f}_\gamma)$, where $\alpha,\beta,\gamma=L,R$ denotes the Fermi distribution function in left (right) leads, as $\mathcal{G}^{\lessgtr}$ contains terms with $f_{L,R}$.
This is consistent with the results in Ref.\cite{sogo_Manybody_2010} and can be seen as the generalization to the nonequilibrium case.
From this we can find that, in general, if we fix the energy of the incident electrons, the voltage bias would still affect $T_q$ by changing the occupation numbers.
This is consistent with the discussion in Sce.\ref{nonper}, showing the specific particle-hole symmetry in charge-$4e$ SC\@.

As for other three types of Andreev processes, the lowest order expansions of $T_{\alpha\beta\gamma}$ can be expressed like (omitting the expressions of $A_{1,2,3}$ for simplification)
\begin{eqnarray}
    &&T^{(1)}_{LLR}
    = 4z^3 v^2\cdot\frac{
	    A_1(\omega^e_1,\omega^e_2,\omega^e_3,\omega^e_4, z_1)
    }{
        B(\omega^e_1,\omega^e_2,\omega^e_3,\omega^e_4)
    } ,
\nonumber\\
    &&T^{(1)}_{LRR}
    = 4z^2 v^2\cdot\frac{
	    A_2(\omega^e_1,\omega^e_2,\omega^e_3,\omega^e_4, z_1)
    }{
        B(\omega^e_1,\omega^e_2,\omega^e_3,\omega^e_4)
    } ,
\nonumber\\
    &&T^{(1)}_{RRR}
    = 4z v^2\cdot\frac{
	    A_3(\omega^e_1,\omega^e_2,\omega^e_3,\omega^e_4, z_1)
    }{
        B(\omega^e_1,\omega^e_2,\omega^e_3,\omega^e_4)
    } ,
\end{eqnarray}
with $v=\Delta/\pi\rho$ and $z_1=\pi^2\rho^2 |t_c|^2/(1+\pi^2\rho^2 |t_c|^2)$.
Despite having different sensitivity to the coupling, the same form of denominator of $T^{(1)}_{\alpha\beta\gamma}$ shows similar energy dependence.
We can write it explicitly
\begin{eqnarray}
    B(\omega^e_1,\omega^e_2,\omega^e_3,\omega^e_4)
    = \prod_{i\ne j} {(\omega^e_i-\omega^e_j)}^2 .
\end{eqnarray}
Therefore, the perturbation breaks down when at least two of the incident electrons carry the same energy.
In fact, it is these points that mainly contributes to $T_{\alpha\beta\gamma}$ and dominants the transmission processes.

\section{The Equation of Motion for Retarded Green's Functions}

In this Appendix, we derive the EOM for charge-$4e$ anomalous Green's function.

\subsection{\label{eom_f}The Equation of Motion for Multi-time Retarded Green's Function}

According to the Eq. (\ref{eq_fr}), We begin with examining the time evolution of multi-time retarded Green's function $\mathscr{F}$.
For convenience of future applications, we consider the general form of $\mathscr{F}(t_x,t_y,t_z)$,
\begin{eqnarray}
    &&\mathscr{F}
    ( A(0), X(t_x), Y(t_y), Z(t_z) )
\nonumber\\
&&\qquad
    = P(xyz)
    \theta(0-t_x) \theta(t_x-t_y) \theta(t_y-t_z)
\nonumber\\
&&\qquad\qquad
    \{[\{ A(0), X(t_x) \}, Y(t_y) ], Z(t_z) \} ,
\end{eqnarray}
with operators $A,X,Y,Z$ owning different times.
The Fourier transform of $\mathscr{F}(t_x,t_y,t_z)$ can be written as
\begin{widetext}
\begin{eqnarray}\label{eq_ft}
    \mathscr{F}
    (\omega_x,\omega_y,\omega_z)
    &&=
    \int \mathrm{d}t_x e^{i\omega_x t_x}
    \int \mathrm{d}t_y e^{i\omega_y t_y}
    \int \mathrm{d}t_z e^{i\omega_z t_z}
    \mathscr{F} (t_x,t_y,t_z)
    \nonumber \\
    &&= \int^0_{-\infty}
    dt_x e^{i\omega_x t_x}
    \int^{t_x}_{-\infty}
    dt_y e^{i\omega_y t_y}
    \int^{t_y}_{-\infty}
    dt_z e^{i\omega_z t_z}
    \mathscr{F} (t_x,t_y,t_z) .
\end{eqnarray}
\end{widetext}
We first consider the time derivative of $t_z$.
Using the Heisenberg equation, we have
\begin{eqnarray}
    i\partial_{z} \mathscr{F} (t_x,t_y,t_z)
    = [\mathscr{F},H_z] (t_x,t_y,t_z) ,
\end{eqnarray}
where $[\mathscr{F},H_z]=\mathscr{F}( A(0), X(t_x), Y(t_y), [Z(t_z),H_{tot}(t_z)] )$.
Notice that
\begin{eqnarray}
    &&\int^{t_y}_{-\infty}
    dt_z e^{i\omega_z t_z}\
    i\partial_{z}
    \mathscr{F} (t_x,t_y,t_z)
    = ie^{i\omega_z^- t_y}
    \mathscr{F} (t_x,t_y,t_y)
\nonumber \\
    &&\qquad
    + \omega^-_z
    \int^{t_y}_{-\infty}
    dt_z e^{i\omega_z t_z}
    \mathscr{F} (t_x,t_y,t_z) .
\end{eqnarray}
The Fourier transform of Eq. (\ref{eq_ft}) then gives
\begin{eqnarray}
    &&\omega_z^- \mathscr{F}
    (\omega_x,\omega_y,\omega_z)
\nonumber\\
    &&\quad
    = -i \mathscr{F}_1
    (\omega_x,\omega_{yz})
    + [\mathscr{F},H_z]
    (\omega_x,\omega_y,\omega_z) .
\end{eqnarray}
Here $\mathscr{F}_1 (\omega_x,\omega_{yz})$ is the Fourier transform of $\mathscr{F} (t_x,t_y,t_y)$ with $\omega_{yz}=\omega_y+\omega_z$.
We note that in our calculations, $\omega_{x,y,z}$ are hole energies due to the selection of time variables that is opposite to the usual way.
This also causes the $-i\eta$ which preserves the causality $0>t_{x,y,z}$.

Repeat the above steps for $t_x$ and $t_y$, we finally get a complete set of EOM,
\begin{eqnarray}
    &&\omega_z^- \mathscr{F}
    (\omega_x,\omega_y,\omega_z)
    = -i \mathscr{F}_1
    (\omega_x,\omega_{yz})
    + [\mathscr{F},H_z]
    (\omega_x,\omega_y,\omega_z) ,
    \nonumber \\
    &&\omega_{yz}^- \mathscr{F}_1
    (\omega_x,\omega_{yz})
    = -i \mathscr{F}_2 (\omega_{xyz})
    + [\mathscr{F}_1,H_y]
    (\omega_x, \omega_{yz}) ,
    \nonumber \\
    &&\omega_{xyz}^- \mathscr{F}_2
    (\omega_{xyz})
    = -i \mathscr{F}_0
    + [\mathscr{F}_2,H_x](\omega_{xyz}) ,
\end{eqnarray}
where $\mathscr{F}_0 = \mathscr{F}( A(0), X(0), Y(0), Z(0))$.
$\mathscr{F}_2 (\omega_{xyz})$, $[\mathscr{F}_1,H_y](\omega_x, \omega_{yz})$ and $[\mathscr{F}_2,H_x](\omega_{xyz})$ are the Fourier transforms of $\mathscr{F} (t_x,t_x,t_x)$, $[\mathscr{F}_1,H_y](t_x, t_y, t_y)$ and $[\mathscr{F}_2,H_x](t_x,t_x,t_x)$ with
\begin{eqnarray}
    &&[\mathscr{F}_1,H_y]
    = \mathscr{F}
    (A(0), X(t_x),[Y(t_y),H_{tot}(t_y)], Z(t_y))
\nonumber \\
    &&\qquad
    + \mathscr{F}
    (A(0), X(t_x),Y(t_y), [Z(t_y),H_{tot}(t_y)]) ,
\nonumber \\
    &&[\mathscr{F}_2,H_x]
    = \mathscr{F}
    (A(0), [X(t_x),H_{tot}(t_x)],Y(t_x), Z(t_x))
\nonumber \\
    &&\qquad
    + \mathscr{F}
    (A(0), X(t_x), [Y(t_x),H_{tot}(t_x)], Z(t_x))
\nonumber \\
    &&\qquad
    + \mathscr{F}
    (A(0), X(t_x) ,Y(t_x), [Z(t_x),H_{tot}(t_x)]) .
\end{eqnarray}

\subsection{\label{eom_eq}The Equation of Motion for Equilibrium Retarded Green's Function}

We here give the explicit form of the EOM to solve $\mathscr{F}(\omega_x,\omega_y,\omega_z)$.
Using the EOM derived in Appendix~\ref{eom_f} and the commutation relations Eq. (\ref{eq_com}), we define $\mathbf{F}$, which is the Fourier transform of
\begin{eqnarray}
    \begin{pmatrix}
    \mathscr{F}(A(0),X_1(t_x),Y_1(t_y),Z_1(t_z)) \\
    \mathscr{F}(A(0),X_1(t_x),Y_1(t_y),Z_2(t_z)) \\
    \mathscr{F}(A(0),X_1(t_x),Y_1(t_y),Z_3(t_z)) \\
    \end{pmatrix} ,
\end{eqnarray}
with $A=\tilde{c}_{1k}, X_1=\tilde{c}_{xk}, Y_1=\tilde{c}_{yk}$ and $Z_1=\tilde{c}_{zk}, Z_2=\tilde{d}^\dagger_{zk}, Z_3=\tilde{\xi}_{zk}\tilde{c}_{zk}$.
This leads to
\begin{eqnarray}
    (\omega^-_z\ \mathbf{I}_{3} - \mathcal{H})
    \mathbf{F}
    = -i \mathbf{F}'_1 ,
\end{eqnarray}
with the coefficient matrix
\begin{eqnarray}
    \mathcal{H}=
    \begin{pmatrix}
    \epsilon_k & \Delta & 0 \\
    0 & -3\epsilon_k & \Delta \\
    0 & \Delta & \epsilon_k \\
    \end{pmatrix} ,
\end{eqnarray}
and $\mathbf{F}'_1$ being the Fourier transform of
\begin{eqnarray}
    \begin{pmatrix}
    \mathscr{F}(A(0),X_1(t_x),Y_1(t_y),Z_1(t_y)) \\
    \mathscr{F}(A(0),X_1(t_x),Y_1(t_y),Z_2(t_y)) \\
    \mathscr{F}(A(0),X_1(t_x),Y_1(t_y),Z_3(t_y)) \\
    \end{pmatrix} .
\end{eqnarray}
It then requires $\mathbf{F}'_1$ to solve $\mathbf{F}$.
We thus define $\mathbf{F}_1$, in which the $(3i+j-3)$ element is the Fourier transform of $\mathscr{F}(A(0),X_1(t_x),Y_i(t_y),Z_j(t_y))$ with $i,j=1,2,3$.
Here the definition of $Y_{1,2,3}$ is the same as $Z_{1,2,3}$ and we can get $\mathbf{F}'_1$ as the top 3 elements of $\mathbf{F}_1$.

Following the similar procedures, we can derive the EOM for $\mathbf{F}_1$,
\begin{eqnarray}
    (\omega^-_{yz}\ \mathbf{I}_{9} - \mathcal{H}_1)
    \mathbf{F}_1
    = -i \mathbf{F}'_2 ,
\end{eqnarray}
with the coefficient matrix
\begin{widetext}
\begin{eqnarray}
    \mathcal{H}_1=
    \begin{pmatrix}
    \mathcal{H}_{11} & \Delta\mathbf{I}_{3} & \mathbf{0} \\
    \mathbf{0} & \mathcal{H}_{12} & \Delta\mathbf{I}_{3} \\
    \mathbf{0} & \Delta\mathbf{I}_{3} & \mathcal{H}_{11} \\
    \end{pmatrix} , \
    \mathcal{H}_{11}=
    \begin{pmatrix}
    2\epsilon_k & \Delta & 0 \\
    0 & -2\epsilon_k & \Delta \\
    0 & \Delta & 2\epsilon_k \\
    \end{pmatrix} , \
    \mathcal{H}_{12}=
    \begin{pmatrix}
    -2\epsilon_k & \Delta & 0 \\
    0 & -6\epsilon_k & \Delta \\
    0 & \Delta & -2\epsilon_k \\
    \end{pmatrix}.
\end{eqnarray}
\end{widetext}
Besides, $\mathbf{F}'_2$ is the top 9 elements in $\mathbf{F}_2$, in which the $(9i+3j+k-12)$ element is the Fourier transform of $\mathscr{F}(A(0),X_i(t_x),Y_j(t_x),Z_k(t_x))$ with $i,j,k=1,2,3$.
Here the definition of $X_{1,2,3}$ is the same as $Z_{1,2,3}$.

We then write the EOM for $\mathbf{F}_2$
\begin{eqnarray}
    (\omega^-_{xyz}\ \mathbf{I}_{27} - \mathcal{H}_2)
    \mathbf{F}_2
    = -i \mathbf{F}_0 ,
\end{eqnarray}
with $\mathcal{H}_2$ being the coefficient matrix similar to $\mathcal{H}, \mathcal{H}_1$.
$\mathbf{F}_0$ is the vector with its $(9i+3j+k-12)$ element being $\mathscr{F}(A(0),X_i(0),Y_j(0),Z_k(0))$.

Sincs the value of $\mathbf{F}_0$ can be solved exactly\cite{li_Charge_2022},
at zero temperature, we find that the nonzero terms in $\mathbf{F}_0$ are
\begin{eqnarray}
    &&\mathscr{F}(A(0),X_2(0),Y_1(0),Z_1(0))=1 ,
    \nonumber\\
    &&\mathscr{F}(A(0),X_2(0),Y_1(0),Z_3(0))=1 ,
    \nonumber\\
    &&\mathscr{F}(A(0),X_3(0),Y_1(0),Z_2(0))=-1 .
\end{eqnarray}
This completes the whole set of EOM and we can solve $\mathscr{F}(\omega_x,\omega_y,\omega_z)$ as $\mathbf{F}(1)$.

\subsection{\label{eom_neq}The Equation of Motion for Nonequilibrium Retarded Green's Function}

Here we extend the EOM in Appendix~\ref{eom_eq} to the nonequilibrium case with the coupling of the left normal metal lead and briefly discuss the complexity for transport in charge-$4e$ SC\@.
In charge-$2e$ SC, the EOM for nonequilibrium Green's functions can still be closed with the help of commutation relations like
\begin{eqnarray}
    [c_{1k\sigma},H_{C}]
    = t_c a_{1\sigma}
     , \quad
    [a_{1k\sigma}, H_{C}]
    = t_c c_{1\sigma} ,
\end{eqnarray}
where $a_{1\sigma} = \sum_k a_{1k\sigma}$ and $c_{1\sigma} = \sum_k c_{1k\sigma}$.
However, the situation for charge-$4e$ SC is much more complex, due to the commutation relations like
\begin{widetext}
\begin{eqnarray}
    &&[d^\dagger_{1k\uparrow},H_{C}]
    = -t_c\ \big(
    a^\dagger_{1\downarrow}
    c^\dagger_{2k\uparrow}
    c^\dagger_{2-k\downarrow}
    +
    c^\dagger_{1-k\downarrow}
    a^\dagger_{2\uparrow}
    c^\dagger_{2-k\downarrow}
    +
    c^\dagger_{1-k\downarrow}
    c^\dagger_{2k\uparrow}
    a^\dagger_{2\downarrow}
    \big) ,
    \label{c17}\\
    &&
    [\xi_{1k\uparrow}c_{1k\uparrow}, H_{C}]
    = t_c \ \xi_{1k\uparrow} a_{1\uparrow}
    + \big(
    j_{1-k\downarrow} \nu_{2k\uparrow, 2-k\downarrow}
    +
    j_{2k\uparrow} \nu_{1-k\downarrow, 2-k\downarrow}
    +
    j_{2-k\downarrow} \nu_{1-k\downarrow, 2k\uparrow}
    \big) c_{1k\uparrow} \label{c18},
\end{eqnarray}
\end{widetext}
where $a^\dagger_{i\sigma} = \sum_k a^\dagger_{ik\sigma} , a_{i\sigma} = \sum_k a_{ik\sigma}$,
$\nu_{ik,jk} = n_{ik} n_{jk} - \bar{n}_{ik}\bar{n}_{jk}$ and $j_{ik\sigma}=t_c\sum_{k'} (c^\dagger_{ik\sigma} a_{ik'\sigma} - a^\dagger_{ik'\sigma} c_{ik\sigma})$.
Here Eq. (\ref{c17}) describes the coupling between charge-$3e$ particles and left leads, which would mix the momentum of three channels,
and Eq. (\ref{c18}) describes the effect that the occupation numbers in charge-$4e$ SC would be affected by the appearance of the current.

To suppress these many-body correlation effects, we consider the approximation with weak coupling $|t_c|^2$, where the lead coupling only introduces an extra self-energy $\mathcal{E}$ to the charge-$4e$ anomalous Green's function.
This approximation is effective due to the dominant contribution of the diagonal term $\delta_{kk_1}\delta_{kk_2}\delta_{kk_3}\delta_{kk_4} F^r_{k_1k_2k_3k_4}$ to the summation $\sum_{k_i} F^r_{k_1k_2k_3k_4}$ as those momentum mixing terms are higher order terms of $|t_c|^2$.
Therefore, we have
\begin{eqnarray}
    (\omega^-_z\ \mathbf{I}_{3} - \mathcal{H} - \mathcal{E})
    \mathbf{F}_{neq}
    &&= -i \mathbf{F}'_{1,neq} ,
\nonumber\\
    (\omega^-_{yz}\ \mathbf{I}_{9} - \mathcal{H}_1 - \mathcal{E}_1)
    \mathbf{F}_{1,neq}
    &&= -i \mathbf{F}'_{2,neq} ,
\nonumber\\
    (\omega^-_{xyz}\ \mathbf{I}_{27} - \mathcal{H}_2  - \mathcal{E}_2)
    \mathbf{F}_{2,neq}
    &&= -i \mathbf{F}_0 ,
\end{eqnarray}
with
\begin{eqnarray}
    &&\mathcal{E}=
    \begin{pmatrix}
    i\Gamma^L/2 & 0 & 0 \\
    0 & 3i\Gamma^L/2 & 0 \\
    0 & 0 & i\Gamma^L/2 \\
    \end{pmatrix} ,
\end{eqnarray}
and
\begin{widetext}
\begin{eqnarray}
    \mathcal{E}_1=
    \begin{pmatrix}
    \mathcal{E}_{11} & 0 & 0 \\
    0 & \mathcal{E}_{12} & 0 \\
    0 & 0 & \mathcal{E}_{11} \\
    \end{pmatrix}
    ,
    \mathcal{E}_{11}=
    \begin{pmatrix}
    i\Gamma^L & 0 & 0 \\
    0 & 2i\Gamma^L & 0 \\
    0 & 0 & i\Gamma^L \\
    \end{pmatrix}
    ,
    \mathcal{E}_{12}=
    \begin{pmatrix}
    2i\Gamma^L & 0 & 0 \\
    0 & 3i\Gamma^L & 0 \\
    0 & 0 & 2i\Gamma^L \\
    \end{pmatrix}.
\end{eqnarray}
\end{widetext}
$\mathcal{E}_2$ is similar to $\mathcal{E}, \mathcal{E}_1$.
Notice that as we consider the coupling hardly altering the property of charge-$4e$ SC, at zero temperature we still use the result of $\mathbf{F}_0$ in equilibrium.
As the coupling being strengthened, not only the contribution of the momentum mixing terms need to be considered but also the value of nonequilibrium $\mathbf{F}_0$ need to be decided by some self-consistent method, which we would leave for future study.

\bibliography{ref1}

\end{document}